\documentclass[prl, aps, floatfix, superscriptaddress, twocolumn]{revtex4-2}

\usepackage{graphicx}
\usepackage{xcolor}

\newcommand{\revision}[1]{{\color{black} #1}} 

\usepackage{bm}
\usepackage{times}
\usepackage{amsmath}
\usepackage{textcomp}

\begin{document}

\title{Magnetically-controlled Vortex Dynamics in a Ferromagnetic Superconductor}

\author{Joseph Alec Wilcox}
\email{Corresponding author: jaw73@bath.ac.uk}
\affiliation{Department of Physics, University of Bath, Claverton Down, Bath, BA2 7AY, United Kingdom}

\author{Lukas Schneider}
\affiliation{Department of Physics, University of Basel, Klingelbergstrasse 82, 4056 Basel, Switzerland}

\author{Estefani Marchiori}
\affiliation{Department of Physics, University of Basel, Klingelbergstrasse 82, 4056 Basel, Switzerland}

\author{Vadim Plastovets}
\affiliation{University of Bordeaux, LOMA UMR-CNRS 5798, F-33405 Talence Cedex, France}

\author{Alexandre Buzdin}
\affiliation{University of Bordeaux, LOMA UMR-CNRS 5798, F-33405 Talence Cedex, France}

\author{Pardis Sahafi}
\affiliation{Department of Physics and Astronomy, University of Waterloo, Waterloo, Canada}
\affiliation{Institute for Quantum Computing, University of Waterloo, Waterloo, Canada}

\author{Andrew Jordan}
\affiliation{Department of Physics and Astronomy, University of Waterloo, Waterloo, Canada}
\affiliation{Institute for Quantum Computing, University of Waterloo, Waterloo, Canada}

\author{Raffi Budakian}
\affiliation{Department of Physics and Astronomy, University of Waterloo, Waterloo, Canada}
\affiliation{Institute for Quantum Computing, University of Waterloo, Waterloo, Canada}

\author{Tong Ren}
\affiliation{Department of Applied Physics, The University of Tokyo, 7-3-1 Hongo, Bunkyo-ku, Tokyo 113-8565, Japan}

\author{Ivan Veshchunov}
\affiliation{Department of Applied Physics, The University of Tokyo, 7-3-1 Hongo, Bunkyo-ku, Tokyo 113-8565, Japan}

\author{Tsuyoshi Tamegai}
\affiliation{Department of Applied Physics, The University of Tokyo, 7-3-1 Hongo, Bunkyo-ku, Tokyo 113-8565, Japan}

\author{Sven Friedemann}
\affiliation{H. H. Wills Physics Laboratory, University of Bristol, Bristol, BS8 1TL, United Kingdom}

\author{Martino Poggio}
\affiliation{Department of Physics, University of Basel, Klingelbergstrasse 82, 4056 Basel, Switzerland}

\author{Simon John Bending}
\affiliation{Department of Physics, University of Bath, Claverton Down, Bath, BA2 7AY, United Kingdom}

\begin{abstract}

    Ferromagnetic superconductors are exceptionally rare because the strong ferromagnetic exchange field usually destroys singlet superconductivity. EuFe$_2$(As$_{1-x}$P$_x$)$_2$, an iron-based superconductor with a maximum critical temperature of 25 K, uniquely exhibits full coexistence with ferromagnetic order below $T_\mathrm{FM}$ \revision{$\simeq$} $19$ K. The interplay leads to narrowing of ferromagnetic domains at higher temperatures and spontaneous nucleation of vortices/antivortices at lower temperatures. Here we demonstrate how the underlying magnetic structure controls the superconducting vortex dynamics in applied magnetic fields. Just below $T_\mathrm{FM}$ we observe a pronounced peak in the creep activation energy, and \revision{magnetic force microscopy measurements reveal the presence of very closely-spaced ($w\ll \lambda$) vortex clusters. We attribute these observations} to the formation of vortex polarons for which we \revision{present} a theoretical description. In contrast, we link strong magnetic irreversibility at low temperatures to a critical current governed by giant flux creep over an activation barrier for vortex-antivortex annihilation near domain walls. Our work suggests new routes for the magnetic enhancement of vortex pinning with important applications in high-current conductors. 

\end{abstract}
\maketitle


\section{Introduction}

The coexistence of ferromagnetism and conventional superconductivity in a single material is extremely rare because the strong ferromagnetic exchange field tends to align the spins of singlet Cooper pairs and destroy them\cite{wolowiecConventionalMagnetic2015}. In the few cases where it has previously been observed, e.g. in rare earth-based rhodium borides\cite{mapleSuperconductivityLongRange1980} and ternary molybdenum chalcogenide Chevrel phases\cite{ishikawaDestructionSuperconductivity1977}, coexistence only occurs over a very narrow range ($\Delta T < 0.5$ K) of rather low temperatures ($T < 1.5$ K) and consists of a spatially modulated magnetic state with a very short period rather than a true ferromagnetic one\cite{monctonOscillatoryMagnetic1980,burletMagnetismSuperconductivity1995}. However, the recent discovery of several europium-containing iron pnictide superconductors has completely transformed this field\cite{micleaEvidenceReentrant2009,anupamSuperconductivityMagnetism2009,liuSuperconductivityFerromagnetism2016,liuNewFerromagnetic2016}. In particular, it has been shown that isovalent P-doping in EuFe$_2$(As$_{1-x}$P$_x$)$_2$ leads to the emergence of a dome (Fig.\ \ref{fig:1}a) of high-temperature superconductivity ($T_c(\mathrm{max}) \simeq 25$ K at $x \simeq 0.2$) associated with the Fe-3d electrons whose critical temperature can significantly exceed the ferromagnetic ordering temperature of the Eu$^{2+}$ spins, $T_\mathrm{FM} \simeq 19$ K\cite{renSuperconductivityInduced2009,caoSuperconductivityFerromagnetism2011}. Phosphorus doping also causes the Eu$^{2+}$ magnetic moments to cant out of their initial antiferromagnetic alignment in the $ab$ plane at $x=0$, tilting them very close to the crystalline $c$-axis at $x \simeq 0.2$, and resulting in a large net out-of-plane ferromagnetic moment\cite{herrero-martinMagneticStructure2009,xiaoMagneticStructure2009,zapfVaryingEu2011,nandiCoexistenceSuperconductivity2014,nandiMagneticStructure2014}. Remarkably, due to the spatial separation of the superconducting electrons in the FeAs layers and the Eu$^{2+}$ magnetic sub-lattice, as well an unusually weak exchange interaction, these superconducting and ferromagnetic phases can coexist over a very broad temperature range ($\Delta T \leq 19$ K)\cite{jeevanInterplayAntiferromagnetism2011,nandiCoexistenceSuperconductivity2014,zapfEuropiumbasedIron2017}. In samples close to optimal doping, this offers a unique opportunity to study the influence of uniaxial ferromagnetic order on the superconducting state as it emerges below $T_\mathrm{FM} \simeq 19$ K. 

In a seminal, low-temperature magnetic force microscopy (MFM) imaging study on EuFe$_2$(As$_{0.79}$P$_{0.21}$)$_2$, Stolyarov \emph{et al}.\cite{stolyarovDomainMeissner2018} revealed the striking, cooperative nature of superconductivity and ferromagnetism in this material. As the temperature was lowered below $T_\mathrm{FM}$, their MFM images resolved a ferromagnetic stripe domain structure emerging in the \emph{Domain Meissner State} (DMS). \revision{In this state} the natural domain width was strongly reduced due to the presence of Meissner screening currents flowing near domain walls. At lower temperatures, a first-order transition to the \emph{Domain Vortex State} (DVS) was identified, \revision{in which} dense arrays of vortices and antivortices \revision{were observed to} spontaneously nucleate \revision{within} the ferromagnetic domains. \revision{Moreover}, the \revision{concurrent} suppression of Meissner screening currents \revision{in this state also} led to \revision{the} abrupt growth of domain widths. The presence of the DMS and DVS as bulk phases in EuFe$_2$(As$_{0.8}$P$_{0.2}$)$_2$ was later confirmed by small angle neutron scattering measurements that also revealed the suppression of the two phases at high magnetic fields\cite{jinBulkDomain2022}. In contrast, a follow-up MFM study on a sample with composition $x=0.25$ and $T_c \approx 18.4$ K $ < T_\mathrm{FM}$ revealed a substantially different local magnetic structure that was attributed to the domination of ferromagnetism over superconductivity for this composition\cite{grebenchukCrossoverFerromagnetic2020}.

Previous MFM works have so far focused on elucidating the subtle ways in which the two electronically-ordered phases interact in the absence of an applied magnetic field\cite{stolyarovDomainMeissner2018,grebenchukCrossoverFerromagnetic2020}. \revision{Thus} the influence of the emerging ferromagnetic order on the dynamics of superconducting vortices in an applied magnetic field remains completely unexplored. A comprehensive understanding of this could underpin important applications in high-performance superconducting tapes and/or wires for operation at very high magnetic fields. Here we combine systematic temperature-dependent magnetisation and magnetic relaxation measurements with nanowire MFM imaging experiments. \revision{We} reveal the vortex dynamics in EuFe$_2$(As$_{1-x}$P$_x$)$_2$ crystals in two different doping regimes; the first with $x \approx 0.21$ close to optimal doping with $T_c > T_\mathrm{FM}$ and the second with $x \approx 0.28$ in the overdoped regime with $T_\mathrm{FM} > T_c$. Remarkably, we find that strong magnetic irreversibility only appears in our samples once \textbf{\emph{both}} ordering phenomena are present, i.e.\ $T < T_c$ \textbf{\emph{and}} $T < T_\mathrm{FM}$, clearly highlighting the cooperative nature of the interaction between them. 

Magnetic relaxation measurements in the DMS phase reveal a pronounced peak in the vortex creep activation energy, more than a factor of two larger than the background value at lower temperatures. We attribute this observation to the formation of a \emph{vortex polaron}, when the widths of \textit{up} and \textit{down} domains are locally perturbed by the presence of a nearby superconducting vortex. MFM images provide further evidence for \revision{an attractive vortex-vortex interaction due to} the distortion of the domain structure by vortex polarons, and we also show how penetrating vortices and antivortices lead to shearing and radical restructuring of the underlying ferromagnetic stripe domains. Note that the vortex field in magnetic superconductors induces a polarisation of the localised magnetic moments resulting in some shrinkage of the vortex diameter\cite{bulaevskiiCoexistenceSuperconductivity1985}. When in motion, such vortices polarise the surrounding moments non-uniformly and re-polarise them; these vortices are termed ``polaron-like'' vortices\cite{bulaevskiiPredictionPolaronlike2012,bulaevskiiPolaronlikeVortices2013}. In our case, the vortex polaron is somewhat different, manifesting as a localised distortion of the domain structure. Additionally, the interaction between the vortex and domain magnetic fields leads to a highly unusual short-range attractive vortex-vortex potential and can even stabilise multi-quantum vortices that would not normally exist. Vortex-vortex attraction has been predicted in hybrid superconductor-ferromagnet superlattices\cite{bespalovClusteringVortex2015}, particularly when the magnetic system exhibits strong spatial dispersion. In some sense, our short-period domain structure acts in a similar way, with the scale of magnetic non-locality corresponding to the domain width. \revision{Other exotic cooperative phenomena are also predicted to appear in superconductor-chiral magnet heterostructures, where the mutual interaction of the two states can give rise to a rich variety of mesoscopic states including clusters and stripes\cite{netoMesoscalePhaseSeparation2022}.} 

As the temperature is lowered into the DVS phase, we see a rapid increase in the magnetic remanence and coercivity, \revision{which we} link to a temperature-dependent critical current density governed by giant flux creep over a thermal activation barrier of $\sim$ 240 K. This observation is reminiscent of earlier ac susceptibility studies of vortex-antivortex dynamics in EuFe$_2$(As$_{1-x}$P$_x$)$_2$, where several thermally activated vortex/antivortex hopping mechanisms were identified as being important\cite{ghigoMicrowaveAnalysis2019,prandoComplexVortexantivortex2022}. However, remagnetisation of the stripe domain structure in the DVS phase explicitly requires vortex-antivortex annihilation at domain walls, and we associate the observed thermally activated behaviour with the existence of a Bean-Livingston barrier for this process\cite{devizorovaTheoryMagnetic2019}.

Our results have important implications for the development of high-current superconducting tapes and wires, which are pivotal in applications such as MRI, maglev, and fusion reactors. Although iron-based superconductors generally exhibit lower critical temperatures when compared to the cuprate family of superconductors, their lower anisotropy and better chemical stability present attractive properties that are well suited to industrial-scale fabrication of high-current superconducting tapes and wires\cite{yaoSuperconductingMaterials2021}. A key engineering challenge is the realisation of materials that can sustain high critical current densities while subject to very high magnetic fields\cite{iwasaCaseStudies2009}, an attribute that is strongly dependent on the material's vortex pinning properties. The high-current performance of a superconductor can typically be enhanced through a wide variety of \textit{extrinsic} modifications, e.g.,\ the introduction of non-magnetic\cite{eleyUniversalLower2017} or magnetic\cite{wimbushEnhancedCritical2010} pinning centres, through high-energy particle irradiation\cite{taenEnhancementCritical2012,taenCriticalCurrent2015} or via the magnetic textures in superconductor-ferromagnet multilayers\cite{palermoTailoredFlux2020}.  Our findings indicate that by careful control of the magnetic domain structure in ferromagnetic superconductors, it should be possible to exploit the \textit{intrinsic} phenomena we observe to significantly enhance vortex pinning over a wide range of temperatures and achieve far superior high magnetic field performance.


\section{Results}

\subsection{Magnetic characterisation}

\begin{figure*}
    \centering
    \includegraphics[width=0.95\textwidth]{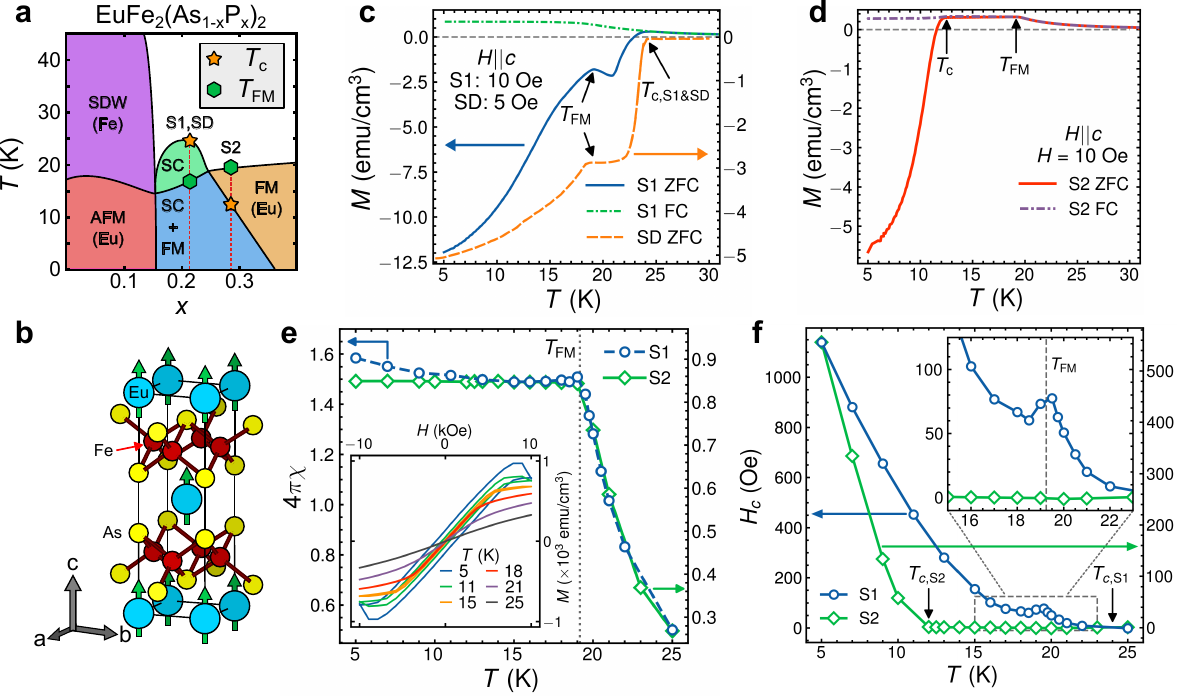}
    \caption{\textbf{Magnetic characterisation of EuFe$_2$(As$_{1-x}$P$_x$)$_2$ samples.} \textbf{a} Schematic phase diagram of EuFe$_2$(As$_{1-x}$P$_x$)$_2$ with approximate positions of samples S1, S2, and SD indicated, after \cite{jeevanInterplayAntiferromagnetism2011,grebenchukCrossoverFerromagnetic2020}. \textbf{b} Crystal structure of EuFe$_2$(As$_{1-x}$P$_x$)$_2$ with direction of Eu moments indicated by green arrows for $x \approx 0.2$. \textbf{c} Zero-field-cooled (ZFC) and field-cooled (FC) measurements of magnetisation for S1 (solid blue and dash-dot green) and SD (dashed orange), in applied magnetic fields of 10 Oe and 5 Oe respectively, oriented parallel to the $c$-axis. Arrows indicate the superconducting critical temperature $T_c$ and ferromagnetic ordering temperature $T_\mathrm{FM}$. \textbf{d} ZFC (solid red) and FC (dash-dot purple) measurements of S2 measured under same conditions as in \textbf{c}. \textbf{e} Temperature dependence of the susceptibility $\chi= \frac{dM}{dH}|_{M=0}$ as determined from MHLs for S1 (blue circles) and S2 (green diamonds). The vertical, dotted grey line indicates $T_\mathrm{FM}$. Inset: example MHLs from S1 at various fixed temperatures. \textbf{f} Coercive field $H_c$ ($M(H_c)=0$) for S1 (blue circles) and S2 (green diamonds), as determined from MHLs. Inset shows expanded view of data in the range 15 K to 23 K to highlight peak in $H_c(T)$ for S1 near $T_\mathrm{FM}$.}
    \label{fig:1}
\end{figure*}

Magnetization data for three single crystals of EuFe$_2$(As$_{1-x}$P$_x$)$_2$ are shown in Figs.\ \ref{fig:1}c and \ref{fig:1}d. Samples S1 and SD both have a doping level close to $x \approx 0.21$ and exhibit identical superconducting onset and ferromagnetic ordering temperatures of $T_c \approx 24.5$ K and $T_\mathrm{FM} \approx 19.3$ K respectively, as shown in the zero-field cooled (ZFC) curves. In contrast, sample S2 with a doping level of $x \approx 0.28$ exhibits the same magnetic ordering temperature of $T_\mathrm{FM} \approx 19.3$ K, but has a much lower superconducting $T_c \approx 12.5$ K. The field-cooled (FC) curves of S1 and S2 are, however, very similar, exhibiting a crossover from paramagnetic to ferromagnetic behaviour at $T_\mathrm{FM}$. This is clearer in measurements with larger applied fields (\textit{Supplementary Information} Fig.\ S1). Given the identified values of $T_c$ and $T_\mathrm{FM}$, the approximate locations of these samples are indicated on the phase diagram shown in Fig.\ \ref{fig:1}a, where S1 and SD correspond to ferromagnetic superconductors ($T_c > T_\mathrm{FM}$) and S2 represents a superconducting ferromagnet ($T_c < T_\mathrm{FM}$).

To characterise the magnetic properties of our samples, families of magnetic hysteresis loops (MHLs) were measured at various fixed temperatures for S1 ($T_c > T_\mathrm{FM}$) and S2 ($T_c < T_\mathrm{FM}$), examples of which are shown for S1 in the inset of Fig.\ \ref{fig:1}e. At the 5 K base temperature, the MHLs of the two samples exhibit features of both superconductivity and ferromagnetism: superconductivity leads to the opening of the hysteresis loop (magnetic irreversibility) and an initial \emph{increase} in the magnitude of the magnetisation upon reversal of the sweep direction at the maximum field excursions, while ferromagnetism is reflected in the steep, linear $M(H)$ behaviour in a window of applied field centred around $H=0$, the width of which increases as the temperature is reduced below $T_\mathrm{FM}$. Above $T_c$ and $T_\mathrm{FM}$ the MHLs of all samples become fully reversible and exhibit a weak, paramagnetic response. 

Fig.\ \ref{fig:1}e illustrates the behaviour of the ferromagnetic contribution to the MHLs as a function of temperature, where $\chi= \frac{dM}{dH}|_{M=0}$ is the local slope where the curves pass through $M=0$. S1 and S2 both display very similar behaviours, showing a rapid increase in $\chi$ as the temperature is reduced from 25 K which saturates in a cusp at the magnetic ordering temperature $T_\mathrm{FM} \approx 19.3$ K, and exhibits only very weak changes at lower temperatures. The temperature at these cusps is very close to those of the features associated with the onset of magnetic order in the ZFC magnetisation curves shown in Figs.\ \ref{fig:1}c and \ref{fig:1}d, and in \textit{Supplementary Information} Fig.\ S1.

The key differences between the two samples become evident in the intermediate temperature regime between 10 K and 20 K. S1 starts to exhibit strong irreversibility below $T_\mathrm{FM} \approx 19.3$ K while S2 remains almost completely reversible until $T < T_c = 12.5$ K. \textit {Evidently the requirement for strongly irreversible behaviour is that both forms of electronic ordering be present.} This is illustrated in the plots of temperature dependent coercive field shown in Fig.\ \ref{fig:1}f. Note that the extremely small coercivity of S2 in the regime $T_c < T < T_\mathrm{FM}$ indicates that the material is an exceptionally soft ferromagnet with very weak domain wall pinning. Similarly \revision{soft} ferromagnetism has also been observed in the end-member of the series, EuFe$_2$P$_2$\cite{fengMagneticOrdering2010}. A more detailed comparison of the reversible component of MHLs for S1 and S2 in this intermediate regime (\textit{Supplementary Information} Fig.\ S2), suggests that S1 is also a soft ferromagnet with very similar properties to S2, and we therefore deduce that any irreversibility $M_\mathrm{irr}$ must be due to the superconductivity. Moreover, the inset to Fig.\ \ref{fig:1}f shows an expanded view of the coercivity for S1 and S2 in the region around $T_\mathrm{FM}$. We see that S1 exhibits a very pronounced coercive field peak in the DMS close to $T_\mathrm{FM}$, something that we attribute to the formation of \textit{vortex polarons}. 

\subsection{Magnetic relaxation measurements}

\begin{figure*}
    \centering
    \includegraphics[width=\textwidth]{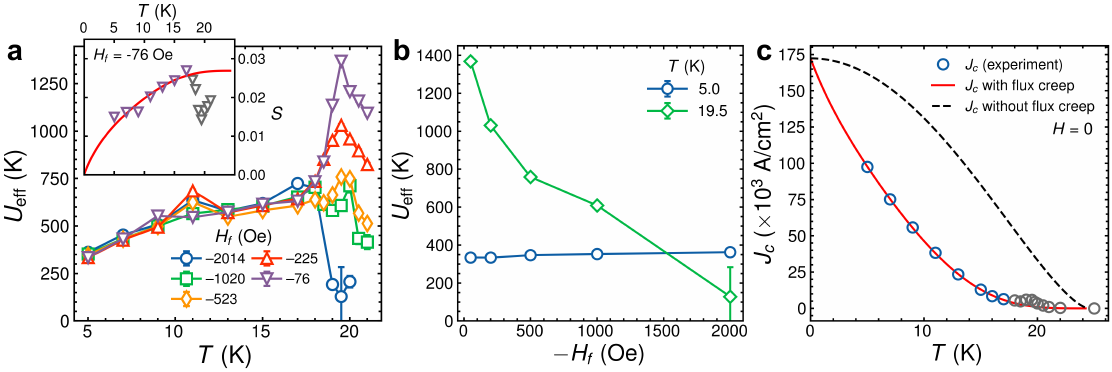}
    \caption{\textbf{Magnetic relaxation and critical current density.} \textbf{a} Effective vortex creep activation energy $U_\mathrm{eff}(T)$ at various final measurement fields $H_f$ for S1. For $H_f$ close to zero, $U_\mathrm{eff}$ exhibits a pronounced peak centred on 19.5 K, which decreases rapidly as the magnitude of $H_f$ increases, before eventually collapsing at $|H_f| \ge 2$ kOe. Inset shows normalised relaxation rate $S(T)$ for $H_f = -76$ Oe. Data for $T \leq 17$ K (purple downward triangles) are fitted to equations \ref{eq:s_thomp} simultaneously with $J_c(T)$ (solid red line) over the same range of $T$. Data above 17 K are not fitted (grey downward triangles). \textbf{b} Effective vortex creep activation energy for S1 as a function of $H_f$ from \textbf{a}, at 5.0 and 19.5 K. $U_\mathrm{eff}$ shows a rapid suppression with $H_f$ at 19.5 K, while the dependence is only very weak at the \revision{lower} temperature. \textbf{c} Critical current density $J_c(T)$ for S1 in the limit of zero applied magnetic field as determined from MHLs. The solid red line is a fit to equation \ref{eq:jc_thomp}, simultaneously with $S(T)$, for data with $T \leq 17$ K (blue circles), while data above are excluded from the fit (grey circles). From the fit we derive a value of $J_c(0) \approx 173$ kA/cm$^2$. The dashed black line is the temperature dependence of $J_c$ in the absence of flux creep.}
    \label{fig:2}
\end{figure*}

To further explore the influence of the underlying ferromagnetism on the superconducting state, particularly in the region of the DMS, we performed magnetic relaxation measurements\cite{yeshurunMagneticRelaxation1996} on sample S1 for $T < T_c$ and for various final measurement fields, $H_f$, after magnetic saturation at $H = +10$ kOe. The time dependence of the irreversible magnetisation, $M_\mathrm{irr}(T)$ was observed to decay logarithmically (\textit{Supplementary Information} Fig.\ S4), from which the normalised relaxation rate, $S(T) = - d\ln{M_\mathrm{irr}/d\ln{t}} = d\ln{J}/d\ln{t}$, was extracted. In the context of the Anderson-Kim model of flux creep\cite{andersonHardSuperconductivity1964}, where the creep activation energy, $U_0$, is linearly reduced by the presence of a bulk current density, the critical current density is expressed by
\begin{equation} \label{eq:jc_ka}
J_c(T) = J_{c0}[1-(T/U_0)\ln{(t/t_\mathrm{eff})}] \, ,
\end{equation}
where $J_{c0}$ is the temperature-dependent critical current density in the absence of flux creep and $t_\mathrm{eff}$ is the effective hopping attempt time. Correspondingly, the normalised relaxation rate, achieved by the logarithmic derivative of equation \ref{eq:jc_ka}, is
\begin{equation} \label{eq:s_ka}
S (T) = - T/[U_0 - T\ln{(t/t_\mathrm{eff})})] \, .
\end{equation}
At low temperatures, the activation energy is well approximated by $U_0 \approx T/|S|$, and it is useful to determine an effective activation energy\cite{yeshurunMagneticRelaxation1996} $U_\mathrm{eff} = T/|S(T,H)|$ to understand the qualitative evolution of the creep activation energy with both temperature and magnetic field.

This is shown for S1 in Fig.\ \ref{fig:2}a, and exhibits two distinct regimes. For $17 \, \mathrm{K} < T < T_{FM}$ the activation energy shows a very pronounced peak centred on 19.5 K ($\sim T_\mathrm{FM}$) with a magnitude more than twice as large than the extrapolated low temperature background at the lowest measurement field. Moreover, this peak rapidly reduces in height as the magnitude of $H_f$ is increased until eventually collapsing towards zero for $-H_f \geq 2$ kOe. In stark contrast, for $T < 17$ K, $U_\mathrm{eff}(T)$ shows a very weak temperature dependence with almost no field dependence up to $-H_f = 2$ kOe, reducing from approximately 600 K to 300 K as the temperature is lowered. The very different behaviour in these two regimes is further emphasised in the plot of $U_\mathrm{eff}(H)$ in Fig.\ \ref{fig:2}b at two characteristic temperatures, and we note that the crossover between the two at $T\approx 17$ K is close to the expected transition between the DMS and DVS phases\cite{stolyarovDomainMeissner2018}.


\subsection{Phenomenological analysis of Abrikosov vortices in ferromagnetic stripe domains}

The presence of a short-period domain structure significantly modifies the properties and mutual interactions of Abrikosov vortices. The origin of the DMS itself lies in the competition between electromagnetic energy, driven by Meissner screening, and the energy associated with magnetic domain walls\cite{daoSizeStripe2011,devizorovaTheoryMagnetic2019}. Since the energy of superconducting vortices is also governed by Meissner screening, a strong interaction between vortices and the magnetic domain structure can be anticipated. A vortex located within one of the domains will, within a characteristic radius $|r|\leq\lambda$, expand neighbouring domains aligned with the orientation of its magnetic moment, and contract those with the opposite orientation, resulting in a deformation of the local domain structure. This interaction leads to the formation of a state we refer to as a \textit{vortex polaron}. 

To provide an insight into the energetics of this scenario, we employ a phenomenological analysis of the free energy of a single Abrikosov vortex sitting within a stripe domain structure with width $l$ and magnetisation oriented along the $z$ direction (see \textit{Supplementary Information} Section S5 for full details of calculation). We find that the energy of a vortex polaron is lower than a standard Abrikosov vortex by an amount
\begin{equation} \label{eq:VP}
    \Delta E = - \frac{\Phi_0^2}{64\pi\lambda l}
\end{equation}
where $\Phi_0$ is the magnetic flux quantum. Thus, there is a substantial lowering in energy if the domain width is smaller than the size of the vortex, as characterised by the penetration depth, $\lambda$. 

The motion of a vortex polaron involves moment reversal near domain walls, resulting in an effective vortex pinning potential and strikingly modified vortex dynamics. Furthermore, the interaction between two vortex polarons can be dramatically modified and the usual repulsive inter-vortex interaction can give way to vortex attraction at short distances smaller than $\lambda$ (but larger than the domain width), favouring vortex clustering. 

\subsection{Giant flux creep in the domain vortex state}

The inset to Fig.\ \ref{fig:2}a shows $S(T)$ at $H_f = -76$ Oe, which is much larger than previously observed in BaFe$_2$(As$_{0.68}$P$_{0.32}$)$_2$ single crystals\cite{salem-suguijrObservationAnomalous2015} but similar in magnitude to other electron-\cite{prozorovVortexPhase2008} and hole-doped\cite{taenCriticalCurrent2015} iron-based superconductors, as well as the \textit{giant flux creep} regime of high-$T_c$ cuprate superconductors\cite{thompsonEffectFlux1993}. The quasi-exponential shape of the critical current density, $J_c(T,H=0)$, shown in Fig.\ \ref{fig:2}c, is also reminiscent of that seen in the cuprates\cite{tamegaiDirectObservation1992,thompsonEffectFlux1993} and iron-based superconductors\cite{taenEnhancementCritical2012,taenCriticalCurrent2015}, suggesting that giant or collective flux creep is important in this material. To describe the behaviour of both $S(T)$ and $J_c(T)$, we base our analysis on a phenomenological model used by Thompson \textit{et al.}\ \cite{thompsonEffectFlux1993} to describe thermally activated flux motion in cuprates, which has also been utilised effectively for similar analysis in iron-based superconductors\cite{taenEnhancementCritical2012}. The authors give the following expressions for $J_c$ and $S$:
\begin{equation} \label{eq:jc_thomp}
    J_c(T) = J_{c0}/[1+(\mu T/ U_0)\ln{(t/t_\mathrm{eff})}]^{1/\mu} \, ,
\end{equation}
\begin{equation} \label{eq:s_thomp}
    S(T) = -T/[U_0 + \mu T \ln{(t/t_\mathrm{eff})}] \, ,
\end{equation}
where $\mu$ is a characteristic, glassy exponent that expresses how $U_0$ depends on the current density. The temperature-dependence of $J_{c0}$ and $U_0$ are assumed to take the forms
\begin{equation} \label{eq:jc0}
    J_{c0}(T) = J_{c00}[1-(T/T_c)^2]^{n_1}
\end{equation}
and 
\begin{equation} \label{eq:u0}
    U_{0}(T) = U_{00}[1-(T/T_c)^2]^{n_2} \, ,
\end{equation}
with $J_{c00} = J_{c0}(0)$ and $U_{00} = U_{0}(0)$. Following Thompson \textit{et al.}\ \cite{thompsonEffectFlux1993}, the exponent $n_1$ is set to be $3/2$, such that $J_{c0}(T) \sim J_\mathrm{depairing}(T)$. However, we allow $n_2$ to be a free fit parameter that reflects the unusual magnetic nature of the creep potential barrier in our samples.

We simultaneously fit $J_c(T, H=0)$ and $S(T, H_f=-76$ Oe) for $T \leq 17$ K to equations \ref{eq:jc_thomp} and \ref{eq:s_thomp} respectively and the results are shown by the solid red lines in the inset of Fig.\ \ref{fig:2}a and Fig.\ \ref{fig:2}c. For the exponent describing the temperature evolution of the activation energy (equation \ref{eq:u0}), we determine a value $n_2 \approx 3$, revealing that $U_0(T)$ is much more rapidly suppressed at high temperatures than when $n = 3/2$ as assumed by Thompson \textit{et al.} This may indicate that the relevant temperature scale of the flux creep mechanism is not $T_c$, but the lower temperature of $T_\mathrm{FM}$. We also determine $U_{00} \approx 235$ K, in very good agreement with the low temperature value of $U_\mathrm{eff}$, and $\mu \approx 1.3$, which is suggestive of vortex-glass\cite{fisherThermalFluctuations1991} or collective-pinning\cite{feigelmanThermalFluctuations1990} scenarios.


\subsection{Magnetic imaging}

\begin{figure*}[!hbt]
    \centering
    \includegraphics[width=\textwidth]{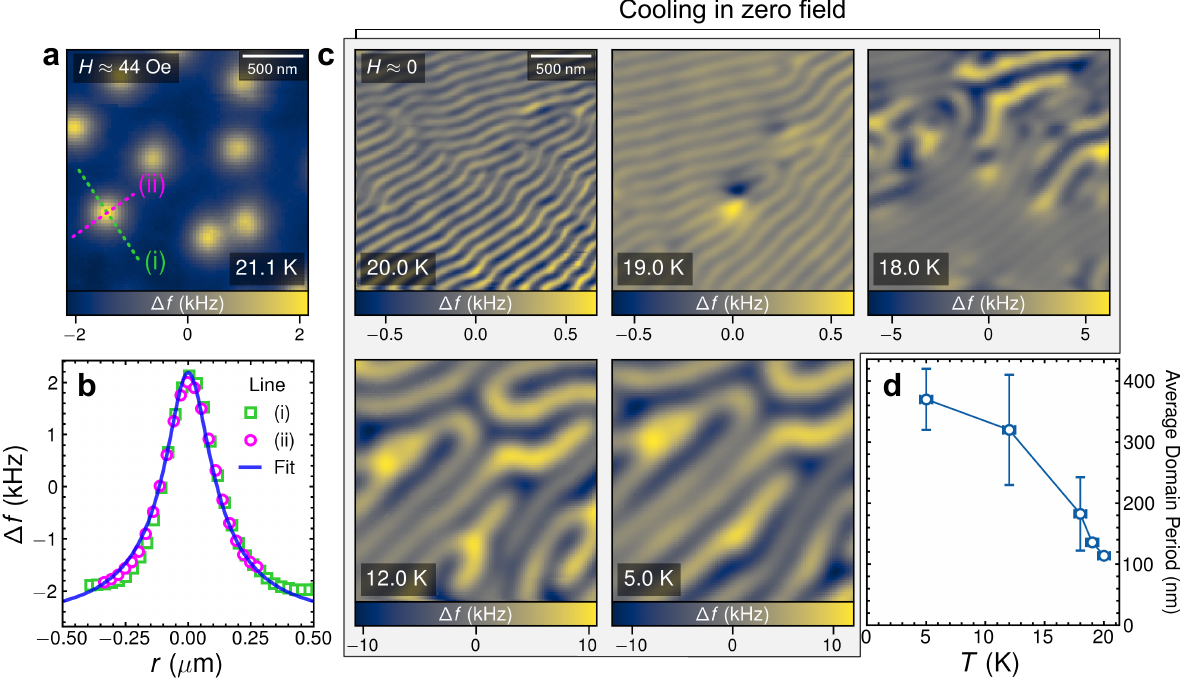}
    \caption{\revision{\textbf{Magnetic texture of sample SD in very low and zero applied field.} \textbf{a} $2\mu\mathrm{m}\times2\mu\mathrm{m}$ scan at $T=21.1$ K, after cooling from above $T_c$ in applied magnetic field $H\approx44$ Oe, showing Abrikosov vortices nucleated on a predominantly flat magnetic background. \textbf{b} Line scans of a vortex in two orthogonal directions, as indicated by the lines (i) [green squares] and (ii) [pink circles] in \textbf{a}. The solid blue line is a fit to a modified variational Clem model. \textbf{c} Series of $2\mu\mathrm{m}\times2\mu\mathrm{m}$ scans at decreasing temperatures in approximately zero applied field. \textbf{d} Average domain period as a function of temperature as determined from MFM measurements in \textbf{c}.}}
    \label{fig:3}
\end{figure*}

To directly visualise how the ferromagnetic state influences the magnetic irreversibility in EuFe$_2$(As$_{1-x}$P$_x$)$_2$, we have undertaken a magnetic force microscopy (MFM) imaging study at a range of different temperatures and magnetic field histories. These measurements were performed using sample SD, which has nominally the same phosphorus composition as S1 and identical values of $T_c$ and $T_\mathrm{FM}$ (Fig.\ \ref{fig:1}c). All images were captured in a plane parallel to the $a$-$b$ surface of the platelet-shaped sample with the field applied along the $c$-axis direction. \revision{Fig.\ \ref{fig:3} shows a series of these MFM images in very low applied magnetic field at several fixed temperatures, where the magnetic contrast is manifest as a shift in the resonant frequency of a nanowire with a ferromagnetic tip (see Methods).}


\begin{figure*}[!hbt]
    \centering
    \includegraphics[width=0.88\textwidth]{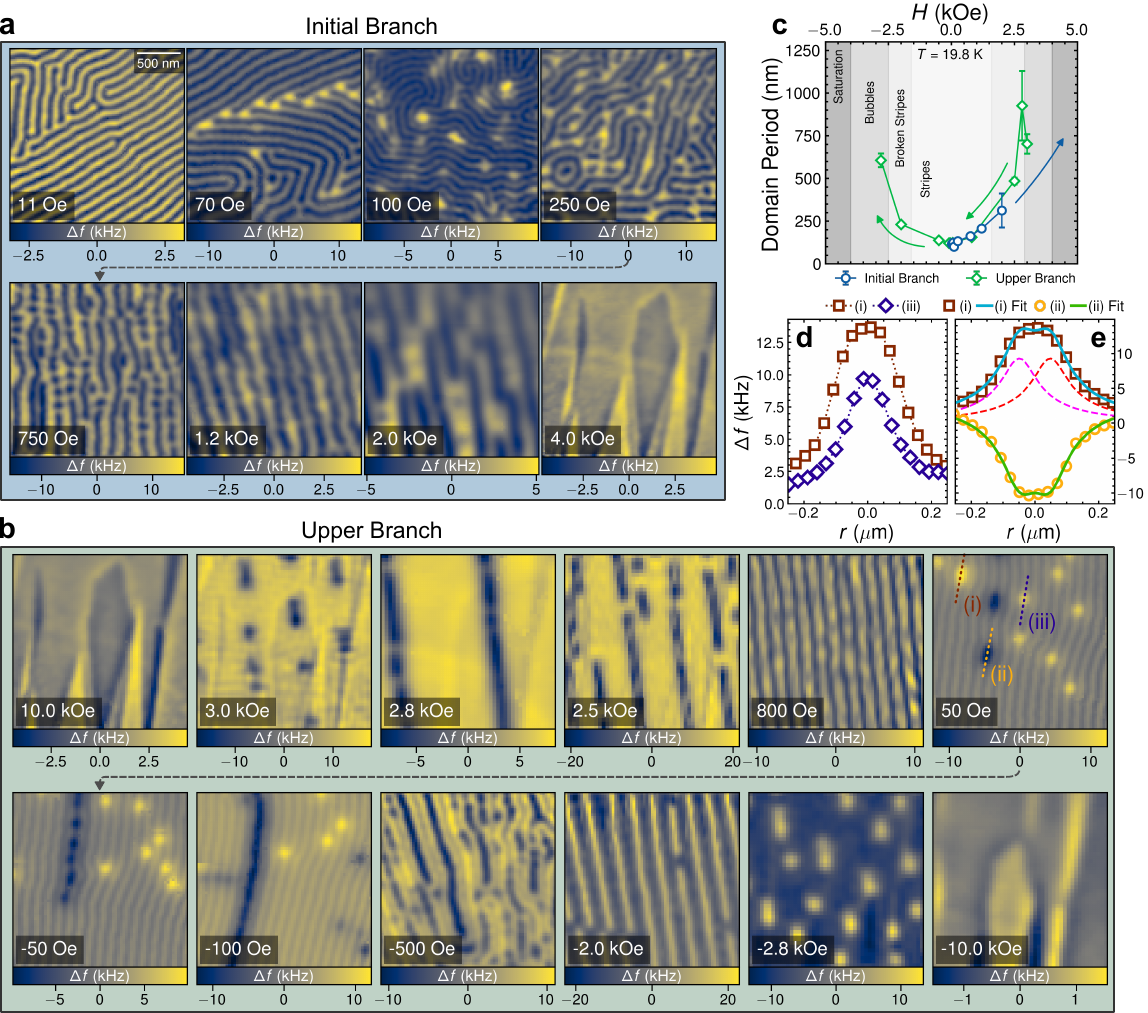}
    \caption{\revision{\textbf{Evolution of the domain Meissner state with applied field.} \textbf{a} and \textbf{b}: Sample SD, series of $2 \mu \mathrm{m} \times 2 \mu \mathrm{m}$ MFM images captured at $T \approx 19.8$ K. \textbf{a} Starting from the ZFC state and increasing the field up to 10 kOe (i.e.\ the initial branch), and \textbf{b} decreasing from 10 kOe, through zero field, and reversing the field to -10 kOe (i.e.\ the upper branch). In \textbf{b}, $H=50$ Oe, three line scans (i) [brown], (ii) [orange] and (iii) [dark blue] are indicated (shown in \textbf{d} and \textbf{e}). \textbf{c} Average domain period as a function of applied magnetic field $H$ from \textbf{a} and \textbf{b} (arrows indicate direction of change of $H$ in the two branches). The shaded areas indicate the approximate magnetic field regions of the various magnetic domain structures observed at this temperature. \textbf{d} Data from line scans (i) [brown squares] and (iii) [dark blue circles] indicate a closely-spaced vortex pair and a single vortex (i.e. $\Delta f > 0$); the former (i) clearly has a larger amplitude and width compared to the latter (iii). \textbf{e} Data from line scans (i) and (ii) [orange circles] showing similarly-sized vortex objects (amplitude and dimension) but with opposite sense (\textit{up} vs.\ \textit{down}). The solid blue and green lines are fits to (i) and (ii) respectively using the bound vortex pair profile, with the dashed red and pink lines indicating the profiles of each vortex in the pair for profile (i).}}
    \label{fig:4}
\end{figure*}

\revision{In Fig.\ \ref{fig:3}a ($T=21.1$ K), the sample is in the purely superconducting state ($T_c > T > T_\mathrm{FM}$), reached by cooling the sample from above $T_c$ in a small residual applied field ($H \approx 44$ Oe). The residual field causes nucleation of Abrikosov flux vortices on a predominantly flat magnetic background. The lines (i) and (ii) indicate line scans across a typical vortex, the profiles of which are shown in Fig.\ \ref{fig:3}b. The MFM system is configured such that the magnetic contrast ($\Delta f$) is proportional to ($\partial B_x/\partial x + \partial B_y/\partial y$) in the $x$-$y$ plane at some height $z$ above the sample. As a result the measured profiles of the vortices, which are proportional to $\partial B_z/\partial z$, appear significantly narrower than if one measured the magnetic field directly. In order to model this shape, we employ the variational Clem model\cite{clemSimpleModelVortex1975} while taking into account the specifics of the contrast mechanism (see \textit{Supplementary Information} Section S6 for further details). The solid blue line in Fig.\ \ref{fig:4}b is a fit of this model to both profiles (i) and (ii), resulting in a very good description of the vortex shape and yielding a scan height $z=83$ nm, consistent with known measurement parameters.} 

\revision{In Fig.\ \ref{fig:3}c, the sample is cooled again from above $T_c$ now in nominally zero applied field, minimising vortex nucleation.} As the sample is cooled below $T_\mathrm{FM}$, a fine stripe domain structure emerges \revision{($T=20.0$ K)}, characteristic of the DMS phase. Light and dark domains have opposite directions of the magnetisation, $\vec{M}$, oriented approximately out of (\textit{up}) and into (\textit{down}) the sample surface. Cooling further to $T=19.0$ K, but remaining above the transition to the DVS, a few spontaneous vortex-antivortex pairs can be seen nucleating around Y-shaped defects in the magnetic domain structure. Vortices(antivortices) appear as much brighter(darker) regions and sit within the \textit{up}(\textit{down}) domains. Further cooling to $T=18.0$ K sees the partial appearance of the DVS, characterised by domains which are much wider (Fig.\ \ref{fig:3}b) and exhibit much stronger magnetic contrast, owing to their high density of spontaneously nucleated vortices and antivortices, and the suppressed Meissner screening currents. The sample, however, does not undergo a uniform transition from the DMS to the DVS as the temperature is reduced due to the first-order nature of the transition\cite{stolyarovDomainMeissner2018, devizorovaTheoryMagnetic2019}, with the DVS component continuing to grow in both domain width and fractional occupation as the temperature is lowered to $T=5.0$ K. \revision{The evolution of the domain size with temperature is shown in Fig.\ \ref{fig:4}d, where the error bars indicate the range in measured domain period of both the DMS and DVS.} The observed zero-field evolution of the DMS and DVS is in good qualitative agreement with previous reports\cite{stolyarovDomainMeissner2018, grebenchukCrossoverFerromagnetic2020}.



\subsection{Field evolution of the domain Meissner state}

Figs.\ \revision{\ref{fig:4}a and b} show a series of MFM images captured in the DMS phase at $T=19.8$ K for a sequence of magnetic fields chosen to recreate the field history of the MHLs of S1 and S2. The sample was first cooled at $H = 0$ to the target temperature, after which the field was increased up to a maximum of $H =10 $ kOe (Fig.\ \ref{fig:4}\revision{a}) building the \textit{initial branch} of the MHL. After ferromagnetic saturation, the field was decreased to zero and then reversed to negative saturation at $H=-10$ kOe, creating the \textit{upper branch} of the MHL (Fig.\ \ref{fig:4}\revision{b}).

The initial branch begins in the pure DMS state, but only a very modest increase of field to $H = 70$ Oe leads to a radical change; a line of \textit{up} vortices penetrating from the sample edge has buckled the stripe domain structure leading to a pronounced cusp-like distortion associated with a line of Y-shaped domain defects. As the field is increased further ($H = 100$ and 250 Oe) this process leads to a complete rearrangement of the domain structure until above 750 Oe the stripes start to align close to the vertical direction, driven by a small in-plane component of the applied field due to an unintentional tilt of the surface normal with respect to the field direction. At the same time, the width of the \textit{up} domains increases with $H$ while the width of the \textit{down} domains decreases, leading to an overall increase in period which is well understood in the context of stripe domain structures in a ferromagnet with uniaxial magnetic anisotropy\cite{hubertMagneticDomains1998}. At high fields, the sample has become penetrated by so much light \textit{up} flux that it is no longer possible to resolve individual vortices, and dark \textit{down} stripes begin to break up into shorter segments and, ultimately, isolated bubbles. Eventually, at $H>4$ kOe the ferromagnet becomes saturated and the domain structure is no longer visible. Any residual contrast in the saturated image is believed to be linked to the surface topography of our samples, with additional contrast arising from the stray field at steps and edges on the sample surface. 

Upon decreasing the field, following the upper MHL branch (Fig. 4\revision{b}), the dark \textit{down} domains reappear via the penetration of magnetic bubbles, presumably containing integer numbers of flux quanta. Further reduction of the field sees these bubbles join up into chains and then fuse into continuous dark stripes. Around zero applied field ($H$ = 50, -50 and -100 Oe), the very short period DMS state is restored, decorated by small numbers of uncorrelated vortices and antivortices. \textit{Up} vortices are confined to \textit{up} domains and vice versa, and all flux structures have slightly elliptical shapes, \revision{predominantly due to the superposition of the domain magnetic fields on the vortex fields.}

Nearly all of the light \textit{up} vortices in the $H$ = 50 Oe image all have the same peak amplitudes and sizes, and are almost certainly single flux quantum vortices. However, one light object in the top-left corner and two dark objects near the centre of the frame have significantly higher amplitudes and are considerably longer. \revision{Additionally, the stripe domain structure appears to become distorted in the vicinity of all flux objects, particularly so for the larger objects. Line profiles of the larger \textit{up} object, one larger \textit{down} object and one smaller \textit{up} object are indicated by the dotted lines (i), (ii) and (iii) respectively, with the profiles shown in Figs.\ \ref{fig:4}d and e. The line scans are taken parallel to the direction of the domains where the contribution of the domain to the profile is approximately constant with position. The profiles of the two \textit{up} objects (Fig.\ \ref{fig:4}d) display clearly distinct characteristics, with (i) being both longer and larger in amplitude than (iii). In contrast, the profiles (i) and (ii), shown in Fig.\ \ref{fig:4}e, display very similar amplitudes and sizes but with opposite polarity, i.e. \textit{up} vs.\ \textit{down}. 

The profiles clearly suggest that the objects in (i) and (ii) are not single flux quantum vortices, as profile (iii) is, but they are pairs of vortices (antivortices) in very close proximity, arranged along the direction of the domain. Using the same model as described for Fig.\ \ref{fig:3}b, but adapted for pairs of vortices (see \textit{Supplementary Information} Section S6), we fit profiles (i) and (ii) with the results shown as the solid blue and green lines in Fig.\ \ref{fig:4}e. Curves have been generated assuming the same fitted scan height of $z = 49$ nm in all cases, and a vortex separation $w = 98$ nm for (i) and $w = 96$ nm for (ii), with the profiles of the individual vortices comprising profile (i) shown as the dashed red and pink lines. Critically, the separation between the vortices is much less than the estimated penetration depth $\lambda(T=19.8\mathrm{K})=462$ nm at this temperature. This pairing of vortices in such close proximity is a key prediction of our \textit{vortex polaron} theory, with the distortion of the domain structure in the vicinity of the vortex objects being another important aspect.}

Following further reduction of the field (Fig.\ \ref{fig:4}e, $H$=-50 and -100 Oe), we first observe a single chain of discrete dark antivortices, occupying the same \textit{down} domain, which then fuses into a structure-less stripe with a very large peak amplitude. Again we believe that this stripe is composed of very closely spaced antivortices held together by \revision{the short-range attraction due to the vortex polaron effect}. As the field is decreased further towards negative saturation, the behaviour mirrors that close to positive saturation except now the light \textit{up} regions become minority domains, shrinking in size and breaking into bubbles. 

\revision{Fig.\ \ref{fig:4}c shows the average domain period determined from this sequence of MFM images, and the approximate regions of the different domain structures are indicated by the shaded areas. In contrast to the magnetisation of sample S1 (e.g.\ $H_c(T)$, Fig.\ \ref{fig:1}f), the domain period shows almost perfect reversibility between the initial and upper branches, indicating the weakness of any domain wall pinning. This further emphasises the fact that the global magnetic irreversibility is driven by superconducting flux dynamics within the ferromagnetic domain structure and not by any intrinsic irreversibility on the part of the ferromagnetism.}


\begin{figure*}[!hbt]
    \centering
    \includegraphics[width=0.75\textwidth]{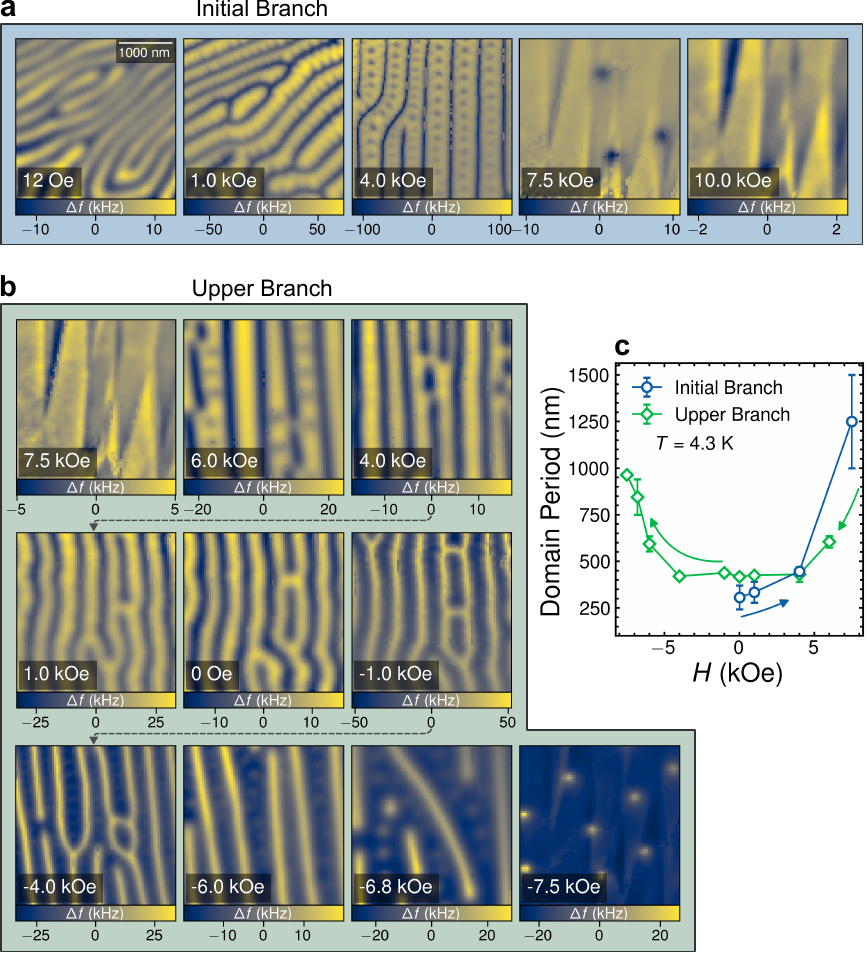}
    \caption{\textbf{Evolution of the domain vortex state with applied field.} \textbf{a} and \textbf{b}: Sample SD, series of $3 \mu \mathrm{m} \times 3 \mu \mathrm{m}$ MFM images captured at $T\approx 4.3$ K, following the same ZFC protocol as in \revision{Figs.\ 4a and b}. \textbf{a} After ZFC and increasing the field to 10 kOe, \textbf{b} decreasing the field from 10 kOe to zero and reversing the field to -10 kOe. \textbf{c} Average domain period as a function of applied magnetic field $H$ from \revision{\textbf{a} and \textbf{b}, with arrows indicating the direction of changing $H$ for the two branches.}}
    \label{fig:5}
\end{figure*}

\subsection{Field evolution of the domain vortex state}

Fig.\ \ref{fig:5} shows a similar series of MFM images to  Fig.\ \ref{fig:4}, but now captured deep in the DVS phase at $T \approx 4.3$ K. While the evolution of the domain structure with field is qualitatively similar, there are several important differences. The domain width \revision{$l$} in the initial ZFC state is now much larger and closer to the intrinsic width of the ferromagnetic domain structure due to the suppression of Meissner screening\cite{devizorovaTheoryMagnetic2019}. Furthermore, these domains are now saturated with a very high density of spontaneously nucleated vortices and antivortices \revision{such that we can no longer resolve discrete vortices in our measurements. Additionally,} any field-induced vortices penetrating the sample \revision{will} experience a magnetic landscape that is markedly different from that in the DMS phase. \revision{We also expect the effects of vortex polaron formation to be substantially weakened in the DVS. This is in part due to the increase of the ratio $l/\lambda$, a key measure of the strength of the effect, but also due to a drastic reduction in the Meissner screening currents in the DVS which are central to the ability of the vortex polaron to reduce the electromagnetic energy of the domain structure.}


When the field is increased from the ZFC state (Fig.\ \ref{fig:5}a), the \textit{up} and \textit{down} domains initially widen and shrink as observed in higher temperature measurements, but the domain structure now survives up to a much higher temperature-dependent saturation field of about $|H|=8$ kOe when the last few magnetic bubbles disappear. Unique to this series of images is the observation of composite domain states of stripes containing chains of bubbles (cf., at $H=4.0$ kOe in Fig.\ \ref{fig:5}a and at $H=6$ kOe in Fig.\ \ref{fig:5}b). While such structures are generally metastable they are often observed in ferromagnets with strong uniaxial anisotropy subject to specific magnetic histories\cite{hubertMagneticDomains1998}. In addition we see a pronounced disordering of the domain structure as the applied field is reduced close to zero with the proliferation of loops linked to Y-shaped defects. There is also noticeable hysteresis in the data\revision{;} the domain period never recovers its initial small ZFC value after the first magnetisation leg \revision{(Fig.\ \ref{fig:5}c)} and magnetic bubbles survive to much higher fields when the applied field magnitude is increasing compared to when it is decreasing. Finally we note that the mechanism by which the sample becomes remagnetised now explicitly involves the penetration of one sign of flux from the sample edges combined with vortex-antivortex annihilation at domain walls. This latter process involves thermal activation over a Bean-Livingston barrier that we attribute as being responsible for the flux creep behaviour observed in magnetisation and magnetic relaxation measurements at lower temperatures.


\section{Discussion}

The magnetometry data for samples S1 and S2 strongly indicate that irreversible vortex dynamics within the ferromagnetic domain structure is the driving force behind magnetic irreversibility in EuFe$_2$(As$_{1-x}$P$_x$)$_2$. Firstly, the purely superconducting state of sample S1 ($T_\mathrm{FM} < T < T_c$) shows very weak magnetic irreversibility, reflecting the absence of Coulomb scattering following isovalent P-doping, as seen in other iron-based superconductors\cite{vanderbeekQuasiparticleScattering2010}. Secondly, the purely ferromagnetic state of sample S2 is highly reversible, and comparison of the reversible magnetisation $M_\mathrm{rev}$ of S1 and S2 establishes the very similar nature of the ferromagnetic ordering at the two different phosphorus compositions. Additionally, the very narrow domain width, apparent in MFM images of sample SD, indicates a very small domain wall energy $\sigma_w$, more than an order of magnitude smaller than e.g.,\ yttrium-iron garnet \cite{guyotDeterminationDomain1973}, that gives rise to very weak domain wall pinning. \revision{This is further emphasised by the high degree of reversibility of the domain size during the MHL shown in Fig.\ \ref{fig:4}}. \revision{Due to the nominally identical characteristics of samples SD and S1}, we argue that \revision{this} must be true for sample S1 as well. Therefore, the rapid increase of magnetic irreversibility when $T < T_c \, \& \, T_\mathrm{FM}$, is clearly a cooperative effect of both superconductivity and ferromagnetism, attributable to the magnetic control of the vortex dynamics. 

The two distinct regimes of the effective vortex pinning potential $U_\mathrm{eff}(T)$ in S1, as well as the peak in $H_c(T)$ near $T_\mathrm{FM}$, clearly indicate a fundamental change in the nature of the magnetically-driven vortex pinning as the sample transitions from the DMS at higher temperatures to the DVS at lower temperatures ($T \lesssim 17$ K). In the high temperature regime, the rapid suppression of $U_\mathrm{eff}(H)$ as the sample is driven to ferromagnetic saturation is an unambiguous signature of the magnetic origin of this behaviour. In the DMS, the magnetisation of the ferromagnetic domains is screened by circulating Meissner currents, causing the domain width to shrink below its intrinsic size\cite{daoSizeStripe2011,devizorovaTheoryMagnetic2019}. Upon transitioning to the DVS, the screening currents collapse in favour of the spontaneous nucleation of vortices and anti-vortices, and the ferromagnetic domains widen back towards their intrinsic values. The behaviour of $U_\mathrm{eff}(T)$ in the two regimes is therefore intimately linked to the spontaneous nucleation of vortices and antivortices as well as the underlying ferromagnetic domain size.

The magnetic irreversibility and magnetic relaxation in S1 in the DMS regime can be understood as being dominated by vortex polaron dynamics. In zero applied field, the domain width is at its narrowest and the vortex polaron energy is at its lowest when compared with a free Abrikosov vortex. The application of, e.g., a positive magnetic field will lead to the penetration of vortices along \textit{up} domains with parallel magnetisation. These domains will widen as the field increases (Fig.\ \ref{fig:4}), leading to a rapid reduction of the energy of vortex polarons associated with them. Therefore, as the sample is driven towards magnetic saturation, the effective pinning potential $U_\mathrm{eff}$ collapses due to the diverging width of the \textit{up} domains and the loss of the associated vortex polaron pinning. In the DVS regime, the domain width is at least \revision{two} time\revision{s} larger than in the DMS state and the vortex polaron energy hence \revision{is} very much lower. In addition the fields of the penetrating free vortices are screened by the surrounding spontaneous vortices and vortex polarons no longer play a significant role.

In contrast, the analysis of $J_c(T)$ and magnetic relaxation $S(T)$ for S1 in the DVS region leads to a consistent picture of giant flux creep with a characteristic activation energy of $U_{00} \approx 240$ K. The mechanism by which the sample becomes remagnetised in the DVS phase must explicitly involve the penetration of one sign of flux from the sample edges in conjunction with vortex-antivortex annihilation at domain walls. The latter process involves thermal activation over a Bean-Livingston barrier\cite{devizorovaTheoryMagnetic2019} which, we believe, governs the flux creep behaviour observed in magnetisation and magnetic relaxation measurements at lower temperatures. We note that thermally activated behaviours with quite similar activation energies have previously been identified in frequency dependent measurements of the ac susceptibility in EuFe$_2$(As$_{0.7}$P$_{0.3}$)$_2$\cite{prandoComplexVortexantivortex2022}. These were attributed to several suggested intra- and inter-domain vortex hopping mechanisms and vortex-antivortex annihilation processes were not explicitly considered.

MFM images at $T=19.8$ K reveal the prolific formation of Y-shaped defects in the domain structure, with a vortex frequently located at the intersection of the domains. These Y-defects appear to be integral to the formation of domain structure grain boundaries (Fig.\ \ref{fig:4}, 70 Oe) and to the observed domain buckling (Fig.\ \ref{fig:4}, 100 and 250 Oe). In these cases we speculate that the dominant vortex penetration direction from the sample perimeter has a large vector component perpendicular to the original domain walls. Since propagation through an adjacent reverse domain has a very large associated energy barrier, it is instead easier for the vortex to distort the stripe domain structure in this direction and travel along the same \textit{up} domain. Recent results from Vagov \textit{et al.} have demonstrated the high mobility of these Y-defects\cite{vagovTemporalEvolution2024} and hence it should be relatively easy for them to stack together, each carrying a vortex at the domain intersection. \revision{Furthermore, the stripe domain structure of the DMS can be viewed as a form of ordered pinning substrate and therefore, with the addition of a suitable drive current, superconducting vortices may exhibit further exotic depinning behaviours and possibly distinct, dynamic phases.\cite{reichhardtDepinningNonequilibriumDynamic2017}}

The vortex polaron potential is inversely proportional to the ferromagnetic domain width (equation \ref{eq:VP}), a parameter that can be tuned by modifying the sample properties. The Kooy-Enz model of ferromagnetic stripe domains\cite{kooyExperimentalTheoretical1960} predicts that the domain period at $H=0$ varies as the square root of the sample thickness, and thinner samples should show narrower domains and a significant enhancement of vortex polaron pinning across a DMS regime that spans a wider range of temperatures. Therefore, manipulation of the domain structure by control of material parameters presents a new route to the magnetic enhancement of vortex pinning strengths in ferromagnetic superconductors that can be employed in the field of high-current tapes/wires for industrial applications.

\section*{Methods}

\subsection{Sample Growth}

Single crystals of EuFe$_2$(As$_{1-x}$P$_x$)$_2$ were grown using a self-flux method. Stoichiometric amounts of FeAs, FeP, and Eu (99.99\%) powders were mixed and loaded into alumina crucibles, which themselves were placed and sealed in stainless steel tubes under Ar atmosphere. The sealed tubes were heated under N$_2$ atmosphere to $\geq 1300^\circ$ C and held for 12 hours, then cooled slowly to $1050^\circ$C at $2^\circ$C per hour before allowing to cool naturally to room temperature. This produces platelet-shaped single crystals with the larger two dimensions corresponding to the $ab$-plane and the shortest dimension to the $c$-axis.

\subsection{Magnetisation Measurements}

A Quantum Design MPMS 3 magnetometer was used to determine $T_c$ and $T_\mathrm{FM}$, and to perform measurements of magnetic hysteresis loops (MHLs) and magnetic relaxation. Measurements were conducted using a quartz half-rod on which a small quartz cube was secured, creating a flat surface the normal of which is parallel to the long axis of the rod and which is located precisely halfway along the half-rod's length. The platelet samples were mounted on this surface so that the crystal $c$-axis was parallel to the axis of the half-rod and thus also to the applied magnetic field. In each measurement, the total magnetic dipole moment $m$ is measured, from which the magnetisation $M$ is derived: $M = m / V$, where $V$ is the volume of the sample. 


Zero-field cooled (ZFC) and field-cooled (FC) measurements were performed by cooling the sample in zero field to base temperature ($\sim$ 5 K), after which a small field was applied and the magnetisation measured upon warming the sample to above $T_c$. The sample is then cooled back to base temperature with the applied field maintained.

MHLs were performed by initially warming the sample above $T_c$ before cooling to the target temperature $T$ in zero field. The magnetisation is measured periodically as the magnetic field is increased to 10 kOe, reduced through zero to -10 kOe and then increased back to 10 kOe. The sweep rate of the magnetic field was kept identical for all MHL measurements, with the same number of measurements within each loop. In complement, a hysteresis loop with increasing field excursions was measured at 5 K (base temperature) in order to determine the minimum field for the establishment of the critical state and full flux penetration of the sample\cite{beanMagnetizationHighField1964,gyorgyAnisotropicCritical1989} (\textit{Supplementary Information} Fig.\ S3). This was found to be $\approx$ 4 kOe, and thus sweeping the field initially to 10 kOe is more than sufficient to achieve full flux penetration. Furthermore, at the same temperature, the full reversal of the critical state was achieved in a window of $\Delta H \approx 1$ kOe. Within the valid critical state portion of each MHL, we calculated the the critical current density $J_c$ using the Bean critical state model for a slab in a perpendicular field\cite{beanMagnetizationHighField1964, thompsonEffectFlux1993}, $J_c(H,T) = 20 \Delta M/(w(1-w/3l))$ (with $J_c$ in units of A/cm$^2$), where $\Delta M = M_\mathrm{upper} - M_\mathrm{lower}$ (in units of emu/cm$^3$) is the width of the hysteresis loop, $l$ is the length of the sample and $w$ is the width of the sample (both in cm), such that $l > w$.

Magnetic relaxation data in sample S1 was taken by warming above $T_c$ and cooling to target $T$ in zero field. The magnetic field is then increased from zero up to 10 kOe, at the same rate as for the MHLs, before decreasing to the final target field $H_f$. Once the final field is reached, the magnetisation was recorded every as a function of time ($M (t)$) every $\sim$ 30 s for several minutes. The time-dependent relaxation of the irreversible magnetisation exhibits a characteristic logarithmic decay and the normalised relaxation rate $S$ is determined from a linear fit to $\ln{M_\mathrm{irr}}$ - $\ln{t}$ (\textit{Supplementary Information} Fig. S4), where $M_\mathrm{irr}$ is the irreversible magnetisation\cite{yeshurunMagneticRelaxation1996}. However, the measurement is of the total magnetisation $M = M_\mathrm{rev} + M_\mathrm{irr}$, where $M_\mathrm{rev}$ is the time-independent reversible contribution to the magnetisation which must be accounted for, using the data from the MHLs, in order to determine the irreversible component only, $M_\mathrm{irr}(t)$.

\subsection{Magnetic Force Microscopy Imaging}

The force microscope used in this study, detailed in references \cite{rossiMagneticForce2019,mattiatNanowireMagnetic2020,marchioriImagingMagnetic2024}, operates with a singly-clamped nanowire as cantilever in the pendulum geometry. The nanowire is made from Si, has a length of 20 $\mu$m and a width of 100 nm. Its fabrication is documented in \cite{sahafiUltralowDissipation2020}. It is tipped with an elongated ferromagnetic Co structure, which renders its two first-order flexural modes susceptible to the magnetic field profile. When modelled with an effective magnetic charge $q$ \cite{mattiatNanowireMagnetic2020, hugQuantitativeMagnetic1998}, the shifts in the mechanical resonance frequency of the two modes, $\Delta f_x$ and $\Delta f_y$, are proportional to the in-plane magnetic field gradients, such that $\Delta f_x = \frac{\partial B_x}{\partial x}$ and $\Delta f_y = \frac{\partial B_y}{\partial y}$. Under the assumption that $\nabla \cdot \mathbf{B} = 0$, the sum of these frequency shifts, $\Delta f = \Delta f_x + \Delta f_y$, is proportional to the out-of-plane field gradient, yielding $\Delta f = -q\frac{\partial B_z}{\partial z}$.

During image acquisition, the nanowire’s high sensitivity to field gradients and the small mode splitting frequently led to mode crossings, preventing the reliable use of a phase-locked loop to track the frequency shifts. Thus, the MFM images were generated by recording thermal noise spectra at each measurement point, extracting the resonance frequencies, $f_x$ and $f_y$, and calculating the frequency shifts according to $\Delta f_x = f_x - f_{x,0}$ and $\Delta f_y = f_y - f_{y,0}$. $f_{x,0} = 243$ kHz and $f_{y,0} = 245$ kHz are the natural resonance frequencies of the modes in the absence of any interaction with the sample. 

MFM was conducted on the platelet-shaped sample SD under varying temperatures and applied magnetic fields. The sample was mounted with its $c$-axis nominally aligned with the field and normal to the imaging plane.  Optical microscopy revealed a small tilt of the $c$-axis with respect to the field, leading to a small in-plane component of the applied field. 

For all measurements the tip-sample separation was between 50 and 150 nm, and was adjusted between scans in order to compensate for the sample-tip interaction strength. For the ZFC measurements, zero applied field was calibrated by minimizing the vortex density to 1-2 vortices per $10 \times 10$ $\mu$m$^2$ area in the purely superconducting state of the sample. The temperature was determined using a four-point probe measurement with a calibrated Cernox\textregistered \ sensor. The sensor is integrated within the heater, which is connected to the sample holder. A small temperature gradient between the sensor and the sample results in a sample temperature which is slightly lower than that read by the sensor, and the magnitude of this difference decreases as the temperature is reduced to the base temperature. The temperatures reported in the manuscript, corresponding to features derived from the MFM images, are therefore the \textit{nominal} temperatures, i.e. the temperatures as read by the sensor. 

\section*{Data Availability}

The data that support the findings of this study are openly available in the University of Bath Research Data Archive at https://(web address will be included here)\cite{wilcoxDatasetMagneticallycontrolledVortex2025}.

\section*{Acknowledgements}

J.A.W.\ and S.J.B.\ acknowledge support from the Engineering and Physical Sciences Research Council (EPSRC) in the United Kingdom under Grant No.\ EP/X015033/1. E.M., L.S.\ and M.P.\ acknowledge support from the Canton Aargau, the Swiss Nanoscience Institute via Ph.D.\ Grant P1905 and the Swiss National Science Foundation via Project Grant No.\ 159893. \revision{E.M., L.S.\ and M.P.\ also acknowledge assistance from the Nano Imaging Lab, Swiss Nanoscience Institute, in the preparation of the nanowire's magnetic tip used in the MFM study.} A.B.\ and V.P.\ acknowledge support by GPR LIGHT and ANR SUPERFAST.

\section*{Author contributions}

J.A.W.\ and S.J.B.\ initiated this work. T.R., I.V.\ and T.T.\ grew the samples. J.A.W., T.R.\ and S.F.\ performed the magnetometry and magnetic relaxation measurements. P.S., A.J.\ and R.B.\ fabricated the ferromagnetic nanowire used in the MFM study. L.S., E.M.\ and M.P.\ performed the MFM measurements. V.P.\ and A.B.\ \revision{developed} the theoretical \revision{model} of the vortex polaron. J.A.W.\ and S.J.B.\ prepared the manuscript with input from all authors.

\section*{Competing interests}

The authors declare no competing interests.

\section*{Additional information}

\textbf{Supplementary Information}
Available online at (URL to be inserted here).\\

\textbf{Correspondence} and requests for materials should be addressed to Joseph A.\ Wilcox or Simon J.\ Bending.

\bibliographystyle{apsrev}
\bibliography{main-text-bibtex}


\end{document}


\title{Supplementary Information for\\``Magnetically-controlled Vortex Dynamics in a Ferromagnetic Superconductor''}

\author{Joseph Alec Wilcox}
\email{Corresponding author: jaw73@bath.ac.uk}
\affiliation{Department of Physics, University of Bath, Claverton Down, Bath, BA2 7AY, United Kingdom}

\author{Lukas Schneider}
\affiliation{Department of Physics, University of Basel, Klingelbergstrasse 82, 4056 Basel, Switzerland}

\author{Estefani Marchiori}
\affiliation{Department of Physics, University of Basel, Klingelbergstrasse 82, 4056 Basel, Switzerland}

\author{Vadim Plastovets}
\affiliation{University of Bordeaux, LOMA UMR-CNRS 5798, F-33405 Talence Cedex, France}

\author{Alexandre Buzdin}
\affiliation{University of Bordeaux, LOMA UMR-CNRS 5798, F-33405 Talence Cedex, France}

\author{Pardis Sahafi}
\affiliation{Department of Physics and Astronomy, University of Waterloo, Waterloo, Canada}
\affiliation{Institute for Quantum Computing, University of Waterloo, Waterloo, Canada}

\author{Andrew Jordan}
\affiliation{Department of Physics and Astronomy, University of Waterloo, Waterloo, Canada}
\affiliation{Institute for Quantum Computing, University of Waterloo, Waterloo, Canada}

\author{Raffi Budakian}
\affiliation{Department of Physics and Astronomy, University of Waterloo, Waterloo, Canada}
\affiliation{Institute for Quantum Computing, University of Waterloo, Waterloo, Canada}

\author{Tong Ren}
\affiliation{Department of Applied Physics, The University of Tokyo, 7-3-1 Hongo, Bunkyo-ku, Tokyo 113-8565, Japan}

\author{Ivan Veshchunov}
\affiliation{Department of Applied Physics, The University of Tokyo, 7-3-1 Hongo, Bunkyo-ku, Tokyo 113-8565, Japan}

\author{Tsuyoshi Tamegai}
\affiliation{Department of Applied Physics, The University of Tokyo, 7-3-1 Hongo, Bunkyo-ku, Tokyo 113-8565, Japan}

\author{Sven Friedemann}
\affiliation{H. H. Wills Physics Laboratory, University of Bristol, Bristol, BS8 1TL, United Kingdom}

\author{Martino Poggio}
\affiliation{Department of Physics, University of Basel, Klingelbergstrasse 82, 4056 Basel, Switzerland}

\author{Simon John Bending}
\affiliation{Department of Physics, University of Bath, Claverton Down, Bath, BA2 7AY, United Kingdom}

\maketitle

\section*{Section S1: Ferromagnetic Transition}

\begin{figure}
    \centering
    \includegraphics[width=0.5\textwidth]{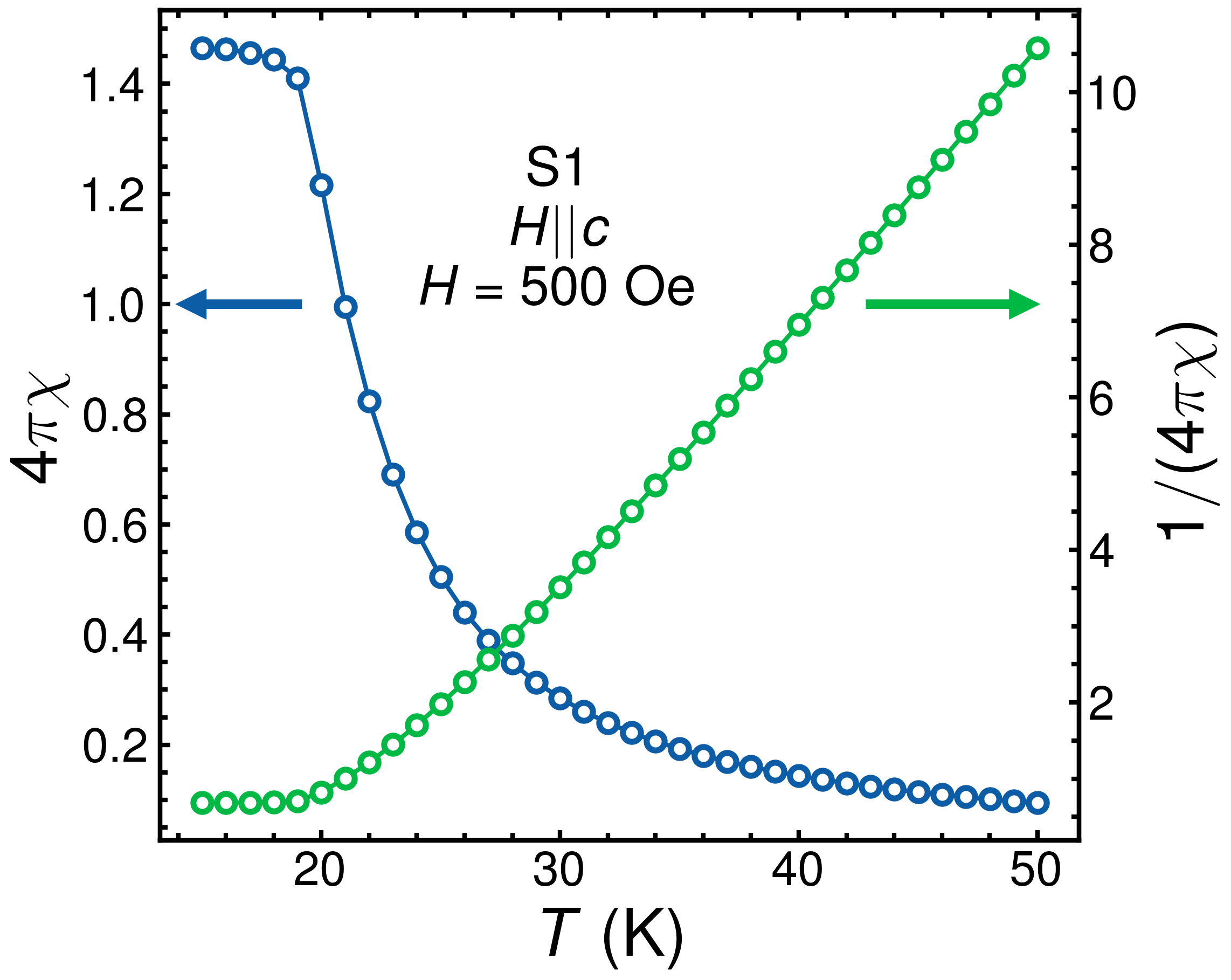}
    \caption{Ferromagnetic transition in sample S1. Field-cooled measurement of the magnetic susceptibility $\chi = M/H$ for EuFe$_2$(As$_{1-x}$P$_x$)$_2$ sample S1 in an applied field of 500 Oe oriented parallel to the crystalline $c$-axis.}
    \label{fig:SM_1}
\end{figure}

In the main text (Fig.\ 1), magnetic characterisation data for EuFe$_2$(As$_{1-x}$P$_x$)$_2$ samples S1 and S2 are presented. In the zero-field cooled curve of S1 the superconducting transition is clearly visible, while the ferromagnetic transition exhibits an unusual form. Figure \ref{fig:SM_1} shows a field cooled measurement of S1 in a larger field, exhibiting a more typical form. The crossover from paramagnetic to ferromagnetic behaviour is evident at $T \approx 19.3$ K, while the superconducting transition at $T\approx 24.5$ K is now no longer visible.

\section*{Section S2: Comparison of Reversible Magnetisation in Samples S1 and S2}

In the main text we discuss the differences in the magnetic characterisation evident in the magnetic hysteresis loops of the two samples S1 ($T_\mathrm{FM} \approx 19.3$ K $< T_c \approx 24.5$ K) and S2 ($T_c \approx 12.5$ K $< T_\mathrm{FM} \approx 19.3$ K). The two samples exhibit similar temperature dependencies of their susceptibilities, but markedly different coercive fields. We argue that the measured magnetic irreversibility is not caused by ferromagnetic domain pinning but by irreversible vortex dynamics that is a result of the presence of both the superconducting and ferromagnetic orders.

\begin{figure}[!hbt]
    \centering
    \includegraphics[width=0.75\textwidth]{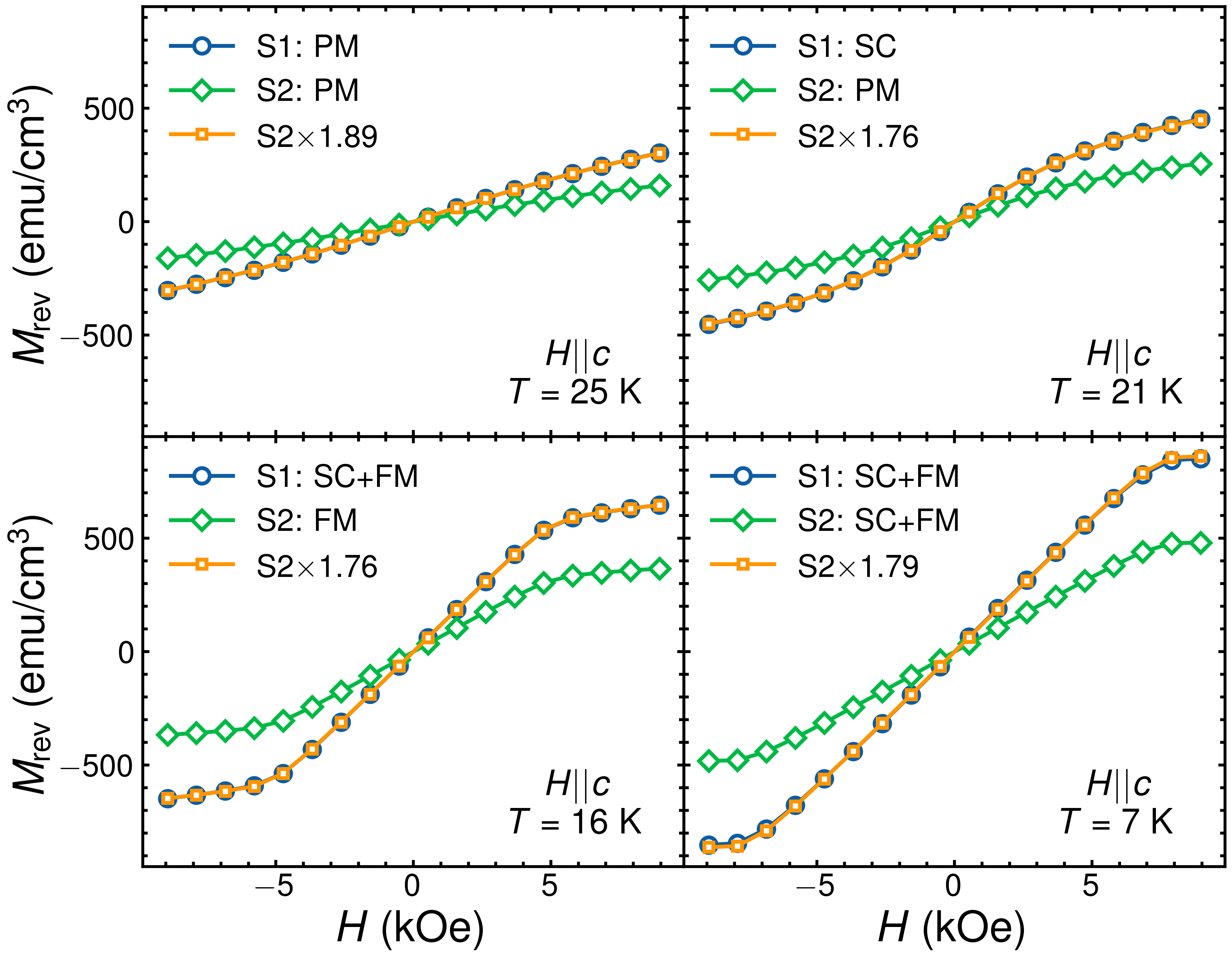}
    \caption{Comparison of the reversible component of the magnetisation $M_\mathrm{rev}(H)$ between samples S1 and S2 at various temperatures. The responses of the two samples scale very well with an $H$-independent, weakly $T$-dependent scale factor.}
    \label{fig:SM_2}
\end{figure}

The reversible component of the magnetisation is $M_\mathrm{rev}(H) = (M_\mathrm{upp}(H) - M_\mathrm{low}(H))/2$, where $M_\mathrm{upp}(H)$ and $M_\mathrm{low}(H)$ are the total magnetisation measured in the upper and lower branches of the hysteresis loop. Figure \ref{fig:SM_2} shows a comparison of $M_\mathrm{rev}$ between samples S1 and S2 at a different temperatures with the two samples in various magnetic phases: paramagnetic (PM), superconducting (SC), ferromagnetic (FM), and simultaneously superconducting and ferromagnetic (SC+FM). The response of either sample can be scaled onto the other with a single field-independent, weakly temperature-dependent scale factor. The scale factor arises in part due to the different sample volumes and also their different aspect ratios, and these two contributions will affect the ferromagnetic and superconducting portion of the sample response differently. 

Given that the two samples exhibit the same reversible magnetic response across the whole temperature range examined, we conclude that there is no significant difference in the paramagnetic and ferromagnetic states of the two samples. Furthermore, in the region where S1 is both SC and FM, and S2 is only FM (i.e. Fig.\ \ref{fig:SM_2}, $T = 16$ K), S2 exhibits a very soft form of ferromagnetism with $H_c \sim 0$ until $T < T_c$ while, contrastingly, $H_c$ in S1 grows quickly as $T$ is reduced (Fig.\ 1f). Since we have argued that the reversible magnetic states of the two samples are very similar, this difference in irreversibility is then precisely due to a combination of the superconducting and ferromagnetic states and not related to the intrinsic, highly-reversible ferromagnetic state that the two samples share.

\section*{Section S3: Establishing the Critical State}

\begin{figure*}[!hbt]
    \centering
    \includegraphics[width=0.9\textwidth]{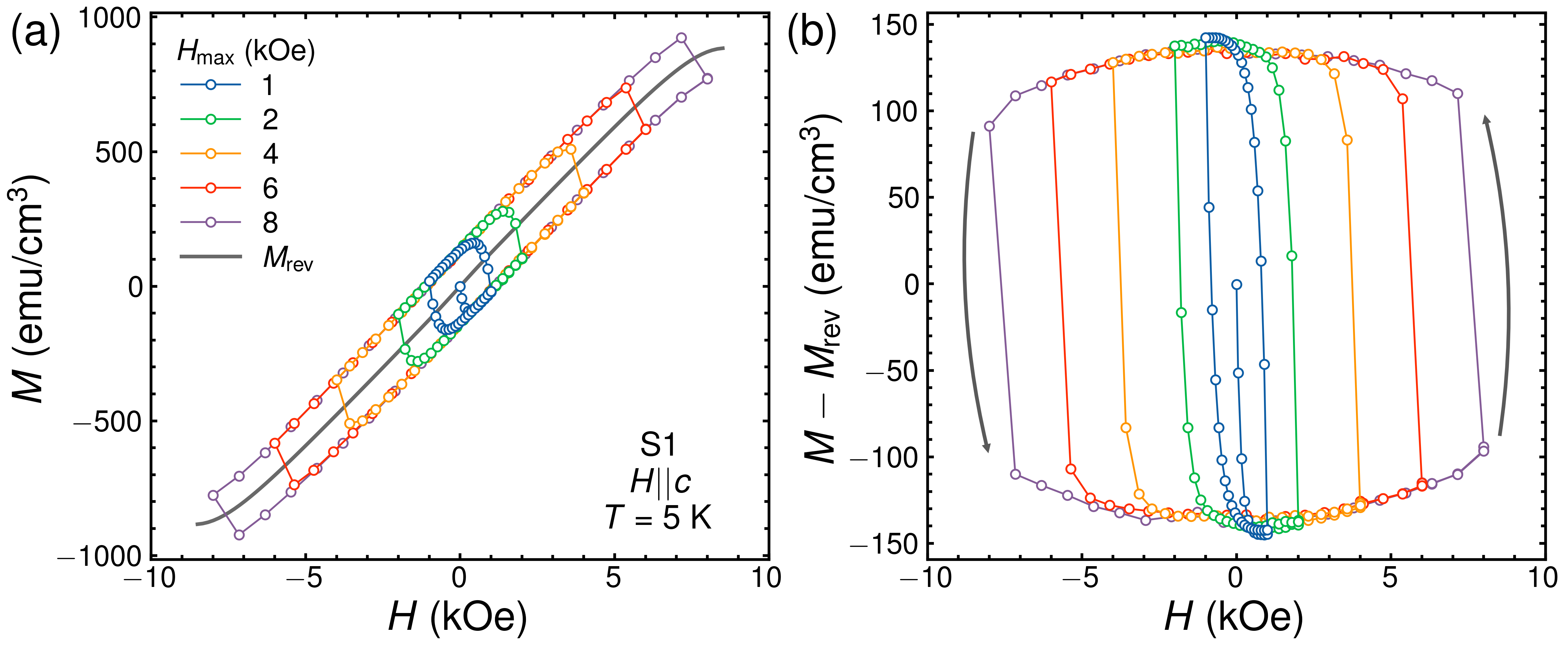}
    \caption{\textbf{Magnetic hysteresis loops with increasing maximum field.} (a) Total magnetisation $M$ of sample S1 measured at $T=5$ K. (b) The irreversible portion of the total magnetisation from the curves in (a). The grey arrows indicate the direction of the change in $M_\mathrm{irr}$ once $H_\mathrm{max}$ is reached and subsequently $|H|$ is reduced.}
    \label{fig:SM_3}
\end{figure*}

In order to make a valid analysis based on the critical state model it is necessary to ensure that the sample has reached full magnetic flux penetration\cite{beanMagnetizationHighField1964,gyorgyAnisotropicCritical1989,eleyUniversalLower2017}. Fig.\ \ref{fig:SM_3}a shows magnetic hysteresis curves measured in sample S1 at $T=5$ K with increasing maximum fields $\pm H_\mathrm{max}$. The initial measurement, $H_\mathrm{max} = 1$ kOe, was made after zero-field cooling the sample from above $T_c$ to 5 K. The field was increased from zero to $+ H_\mathrm{max}$, then reversed to $- H_\mathrm{max}$, and finally increased back to $H_\mathrm{max}$ once more. The subsequent loops were made by continuing the cycle of $+H_\mathrm{max}$ to $-H_\mathrm{max}$ and back again, using the new, larger value of $H_\mathrm{max}$. Additionally, no further repeats of the zero-field cooled process were made. In doing so, we can see how the irreversible state evolves with $H_\mathrm{max}$ and at what point full flux penetration is achieved.

In Fig.\ \ref{fig:SM_3}a, the initial diamagnetic response of the sample can be seen in the first loop, though this fairly quickly gives over to the ferromagnetic response as $H_{p}$, the field of first flux penetration, is exceeded. After the completion of the first curve, all the subsequent curves to increasingly larger values of $H_\mathrm{max}$ can be seen to trace out approximately the same upper and lower branches that are centred on a shared reversible magnetisation $M_\mathrm{rev}$. As noted in the main text, the form of the magnetic irreversibility is superconducting in nature which is evident due to the jump in $M$ at the point of field-sweep reversal at $\pm H_\mathrm{max}$\cite{iwasaCaseStudies2009}. This contrasts to ferromagnetic irreversibility where the magnitude of $M$ does not increase under the same circumstances, only remaining or decreasing smoothly from its maximum achieved at $\pm H_\mathrm{max}$\cite{dellatorreMagneticHysteresis1999}.

Fig.\ \ref{fig:SM_3}b shows the isolated, irreversible portion of the magnetisation $M_\mathrm{irr} = M - M_\mathrm{rev}$ from the hysteresis loops in Fig.\ \ref{fig:SM_3}a. On closer inspection, while there is good overlap of the loops starting from the initial measurement with $H_\mathrm{max} = 1$ kOe, there is a slight discrepancy around $H=0$. It is not until the curve with $H_\mathrm{max} = 4$ kOe that near-perfect overlap with all subsequent curves (i.e. 6 and 8 kOe) is established, indicating that the critical state is fully realised within the sample. Therefore, the field excursions of $\pm 10$ kOe in the magnetic hysteresis loops presented in Fig.\ 1e are more than sufficient to establish the critical state, leading to the analysis and presentation of $J_c(T)$ in Fig.\ 2c. Finally, in Fig.\ \ref{fig:SM_3}b, we also see the clear jumps in magnetisation at the extremes of the loops, as emphasised by the grey arrows. This reversal of the critical state requires a finite window $\Delta H \lesssim 1$ kOe to fully complete, and therefore the calculation of $J_c \propto \Delta M$ should also be avoided in this region (see main text Methods).

\section*{Section S4: Example Magnetic Relaxation Measurements}

\begin{figure}[!hbt]
    \centering
    \includegraphics[width=\textwidth]{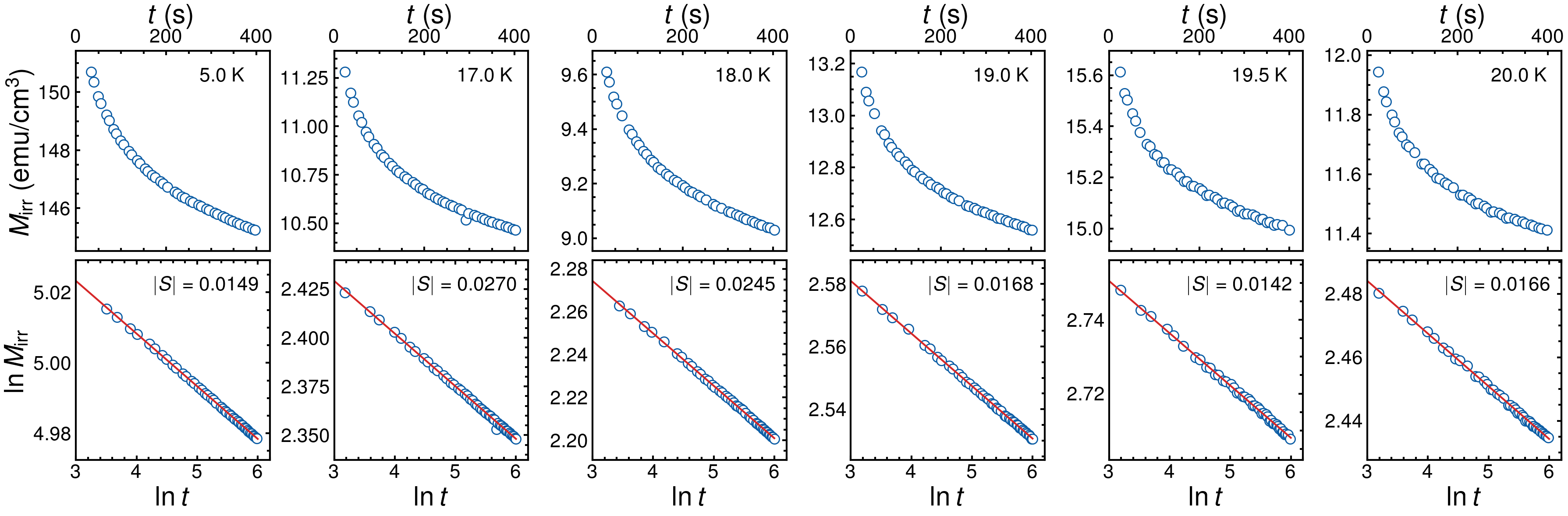}
    \caption{\textbf{Example relaxation measurement data.} Relaxation measurement data for sample S1 measured in a final field $H_f = -76$ Oe, taken at a selection of temperatures. Left column: the irreversible component of the total magnetisation, $M_\mathrm{irr}(t)$, shows a characteristic logarithmic decay. Right column: the normalised relaxation rate, $S$, is determined as the slope of a linear fit to $\ln{M_\mathrm{irr}}$ vs. $\ln{t}$.}
    \label{fig:SM_4}
\end{figure}

Figure \ref{fig:SM_4} shows example magnetic relaxation data taken in a final measurement field of $H_f = -76$ Oe. The selection of temperatures includes the very lowest temperature measured as well as the range of temperature for which anomalous $S(T)$ behaviour was observed (Fig.\ 2,  attributed to vortex polaron dynamics). The time dependence of the irreversible magnetisation, $M_\mathrm{irr}(t)$, displays a very clear, characteristic logarithmic decay for all temperatures. The normalised relaxation rate, $S$, is determined as the slope of a linear fit to $\ln{M_\mathrm{irr}}$ vs $\ln{t}$\cite{yeshurunMagneticRelaxation1996}.

\section*{Section S5: Theoretical model for vortex polarons in ferromagnetic superconductors}

Here we present a theoretical model of the vortex polaron (VP) formation in the domains of a ferromagnetic superconductor. We give a qualitative explanation of the physical origin of the VP, followed by the corresponding calculations. 

\subsection{A. Summary}

\begin{figure}
    \centering
    \includegraphics[width=\linewidth]{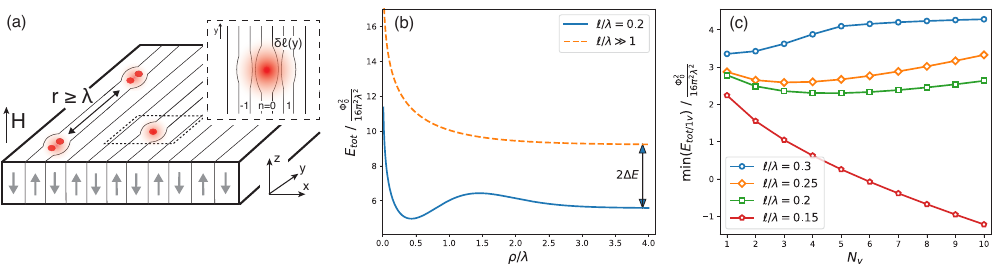}
    \caption{\textbf{Vortex polaron model and calculations.} (a) Sketch of the vortex polaron formation and example of the vortex polaron clustering. (b) Energy of two interacting vortices $E_{tot}(\rho)$ [Eq. (\ref{E_two_v})] with ($\ell/\lambda=0.2$) and without ($\ell/\lambda\gg 1$) polaron effect. (c) Bound energy per one vortex min$E_{tot}$ [Eq. (\ref{E_chain})] in the isolated cluster of $N_v$ vortices. For both plots $\xi/\lambda=0.01$.}
    \label{fig:SI_5}
\end{figure}

The system we study is a superconducting ferromagnet with a domain wall (DW) structure of period $\ell$. Suppose that an applied external magnetic field induces a single Abrikosov vortex with a size of the  order of the London penetration depth $\lambda$ in a domain of the same orientation (see Fig.\ \ref{fig:SI_5}a). If the domain width is smaller than $\lambda$, which is the case at $T\lesssim T_\mathrm{SC}$ (Fig.\ 3a), then the vortex magnetic field will locally expand the hosting magnetic domain. This perturbation, in turn, reduces the electromagnetic energy of the entire system with the only controlling parameter $\ell/\lambda$. In fact, this is a prerequisite for the emergence of a vortex-generated magnetic polaron effect. The VPs can move along the domain and notably interact with each other. As we will show, the magnetic domain is able to mediate the long-range attraction between the vortices, giving rise to the molecule-like few-VPs clusters ($N_v\sim2-3$) with the inner size $\ell\lesssim \rho_0 \lesssim \lambda$. With the decrease of the domain size $\ell$ the attraction strength grows and multi-VP clusters ($N_v\gg1$) can appear, resembling the vortex ``bubbles" in the experimental images (see Figs. 4d and e) due to their small inner size.

\subsection{B. Physical model}

Consider a ferromagnet film with the equilibrium domain structure ${\bf M}_0(x)=\pm \bar{M}_0{\bf z}_0$ ($\bar{M}_0 = \text{const}$) with the period $\ell$. The thickness of the film is assumed to be large enough ($d_\mathrm{F}>\lambda$) so that one can neglect any stray fields and treat the system as homogeneous in the $z$-direction. Thus, the corresponding magnetic field generated by the domains reads as\cite{faureDomainStructure2005}

\begin{gather}
    {\bf B}_0(x)=\frac{16\pi \bar{M}_0}{\ell}\sum_k \frac{q}{q^2+\lambda^{-2}}\sin(q(x+\ell/2)){\bf z}_0  ,
\end{gather}
where $q=(2k+1)\pi/\ell$ and $k$ is integer. The Abrikosov vortex is situated at the origin and has a magnetic field distribution ${\bf B}_v({\bf r})=(\Phi_0/2\pi\lambda^2)K_0(|{\bf r}|/\lambda)$, where $\Phi_0=hc/2e$ is the magnetic flux quantum\cite{tinkhamIntroductionSuperconductivity2015}. Within the framework of the London approximation the total electromagnetic energy of the system per unit length $L_z$ is
\begin{gather}\label{F}
    F = \frac{1}{8\pi}\int d^2r \Big( {\bf B}-4\pi {\bf M} \Big)^2 +\lambda^2 \Big( \nabla\times({\bf B}-4\pi {\bf M}) \Big)^2.
\end{gather}
The appearance of the polaron effect can be obtained within the perturbation approach. Let us introduce a local distortion of the domain created by the vortex as ${\bf M}_0({\bf r})+{\bf M}_1({\bf r})$, assuming $\text{div}{\bf M}_1({\bf r})=0$. The perturbation of the magnetic field ${\bf B}_1({\bf r})$ should be found independently using the linear London equation ${\nabla^2 \big( {\bf B}_1-4\pi {\bf M}_1 \big) = \lambda^{-2}{\bf B}_1}$. Corresponding change in the free energy caused by domain expansion simply reads as
\begin{gather}\label{d_F}
    \Delta E = \frac{1}{4\pi}\int d^2r \ {\bf H}_1 ({\bf H}_0 +  {\bf B}_v) +  \lambda^2  (\nabla\times{\bf H}_1)(\nabla\times({\bf H}_0 +  {\bf B}_v)) +   {\bf H}_1^2+ \lambda^2  (\nabla\times{\bf H}_1)^2 . 
\end{gather}
For the illustrative purposes we will use a simple phenomenological model of the vortex-domain interaction, which will provide us with the intuitive results and estimations. We determine the deformation of the domain wall as  
\begin{equation}\label{M_1}
\begin{split}
    M_1(x,y) = & \bar{M}_0\sum_{n}\text{sgn}\left(x_n-\ell/2\right)-\text{sgn}\left(x_n-\ell/2-\delta\ell(y,n)\right)\\
    & +\text{sgn}\left(x_n+\ell/2+\delta\ell(y,n)\right)-\text{sgn}\left(x_n+\ell/2\right), 
\end{split}
\end{equation}
where $x_n=x-2n\ell$, and $n=0,\pm1,\dots$ is the number of the domain co-directed with the vortex field, starting from the vortex position. The domain profile in the $y$-direction is approximated by the Gaussian function 
\begin{gather}
    \delta \ell(y,n) = \delta\ell_0 ~ \text{exp}\left(-\frac{y^2}{\lambda^2-(2n\ell)^2}\right)~f_n.   
\end{gather}
The amplitude of the deformation of the n-th domain $f_n$ can be connected to the vortex field as $f_n\approx K_0\big(2|n|\ell/\lambda\big)/K_0(\xi)$ with a standard truncation at $\xi$. Here $\delta\ell_0$ is a variational parameter of the problem. The naturally emerged parameter $\ell/\lambda$ determines both the intensity and the spatial distribution of the DW deformation. In order to find the solution of the London equation with ${\bf M}_1$ we utilize the adiabatic approximation taking into account the slow $y$-dependence of the fields, which is justified by the condition $\delta\ell\ll\lambda$. This gives:  
\begin{equation}\label{B_1}
\begin{split}
    B_1(x,y) & = 4\pi \bar{M}_0\sum_{n}
        \text{sgn}\Big(x_n-\ell/2\Big)e^{-\frac{|x_n-\ell/2|}{\lambda}}
        -\text{sgn}\Big(x_n-\ell/2-\delta\ell(y,n)\Big)e^{-\frac{|x_n-\ell/2-\delta\ell(y,n)|}{\lambda}} \\ 
        & \, +\text{sgn}\Big(x_n+\ell/2+\delta\ell(y,n)\Big)e^{-\frac{|x_n+\ell/2+\delta\ell(y,n)|}{\lambda}}
        -\text{sgn}\Big(x_n+\ell/2\Big)e^{-\frac{|x_n+\ell/2|}{\lambda}},
\end{split}
\end{equation}
and consequently $H_1(x,y)=B_1(x,y)-4\pi M_1(x,y)$.

\subsection{C. Vortex polaron energy}

The energy decrease associated with the DWs distortion (\ref{d_F}) can be calculated straightforwardly using Eqs. (\ref{M_1}-\ref{B_1}). To facilitate this step we assume $\ell\lesssim\lambda$ and build up a perturbation theory using the length scale ratio $\ell/\lambda$ as a small parameter. After some derivation we obtain the function $\Delta E(\delta\ell_0)$, with the minimal (optimal) value 
\begin{gather}\label{d_F_0}
    \Delta E  = -\frac{ \Phi_0^2}{32\pi\lambda^2}\frac{C_2^2}{C_1}, 
\end{gather}
\begin{gather}  \notag
    \text{where we have defined} \quad \quad
    C_1 = \sqrt{\frac{\pi}{2}}\sum_{n=0}^{\lambda/2\ell} f_n^2 \sqrt{1-(2n\ell/\lambda)^2} \quad \quad \text{and} \quad \quad
    C_2 = \sum_{n=0}^{\lambda/2\ell} f_n e^{-2n\ell/\lambda}. 
\end{gather}
Note that $C_2^2/C_1$ can be roughly estimated as the number of the domain walls on the scale of the vortex, e.g. $C_2^2/C_1\approx\lambda/2\ell$.
Thus, the correction (\ref{d_F_0}) renormalizes the single vortex energy $E_v\propto \ln(\lambda/\xi)$ as 
\begin{gather}\label{E_v}
    E_\text{VP} = \frac{\Phi
    _{0}^{2}}{16\pi ^{2}\lambda ^{2}} \Bigg[ \ln \left( \frac{\lambda }{\xi }\right) -\frac{ \pi}{2}\frac{C_2^2}{C_1}\Bigg].
\end{gather}
which we refer as the vortex polaron energy. As we already mentioned, the strength of the polaron effect is determined in fact only by the ratio $\ell/\lambda$. One may notice that there is a critical regime at roughly $\ell  <  \pi\lambda/4\ln(\lambda/\xi)$, where the VP energy becomes negative, which means the possibility of VP self-generation. We do not discuss this regime since this was not observed experimentally (see Fig.\ 4), and instead focus on the VP clustering. 

\subsection{D. Interaction of two vortex polarons}

Let us now examine the system of two interacting vortices situated at the distance $\rho$ in the same domain (Fig.\ \ref{fig:SI_5}a). The mutual perturbation of the domain walls $M_1(x,y)=M_1^{(1)}(x,y)+M_2^{(2)}(x,y+\rho)$ creates an effective force between the vortices, which can be extracted directly from Eq. (\ref{F}) using the phenomenological model (\ref{M_1}-\ref{B_1}). The corresponding potential
\begin{multline}
    \Delta E_{int}(\rho) =
    4\Delta E \frac1C_2\sum_{n=0}^{\lambda/2\ell}  f_n e^{-2n\ell/\lambda} e^{-\frac{\rho^2}{\lambda^2(1-(2n\ell/\lambda)^2)}}\\ 
    -2\Delta E\frac{\sqrt{\pi/2}}{C_1}\sum_{n=0}^{\lambda/2\ell}  f_n^2\sqrt{1-\left(2n\ell /\lambda \right)^2}e^{ -\frac{\rho^2}{2\lambda^2(1-(2n\ell/\lambda)^2)}}.
\end{multline}
alters the standard vortex-vortex repulsion potential $E_{v-v}$, and the total energy of two VP becomes:
\begin{equation}\label{E_two_v}
\begin{split}
    E_{tot}(\rho) & = 2E_\text{VP} + E_{v-v}(\rho) + \Delta E_{int}(\rho)\\ 
    & =  \frac{\Phi_{0}^{2}}{8\pi ^{2}\lambda ^{2}}\left[\ln \left( \frac{\lambda }{\xi }\right)  +K_0 \left( \frac{\rho }{\lambda }\right)\right] - 2 \frac{ \Phi_0^2}{32\pi\lambda^2}\frac{C_2^2}{C_1}\times \\ 
    & \left[ 1+2\frac1C_2\sum_{n=0}^{\lambda/2\ell}  f_n e^{-2n\ell/\lambda} e^{-\frac{\rho^2}{\lambda^2(1-(2n\ell/\lambda)^2)}} -\frac{\sqrt{\pi/2}}{C_1}\sum_{n=0}^{\lambda/2\ell}  f_n^2\sqrt{1-(2n\ell/\lambda )^2}e^{ -\frac{\rho^2}{2\lambda^2(1-(2n\ell/\lambda)^2)}} \right].
\end{split}
\end{equation}
The profile of the function (\ref{E_two_v}) is shown in Fig.\ \ref{fig:SI_5}b. One can easily estimate the equilibrium distance between the vortices at $\ell\lesssim\lambda$ as
\begin{gather}\
    \rho_0 \approx  \lambda\sqrt{\frac{2}{3\pi}\frac{C_1}{C_2^2}}\approx \sqrt{\frac{4\lambda\ell}{3\pi}}.
\end{gather}
This means that for $\ell\lesssim\lambda$ the inter-VP distance is $\rho_0\ll\lambda$, what makes this bound (molecule-like) structure almost indistinguishable from a two-quanta vortex. 
We note that VPs from different domains can also interact in the $x$-direction. This question, however, appears to be less relevant and is therefore omitted here. 

\subsection{E. Chain of vortex polarons}

Now let us consider a finite chain of vortex polarons $N_v$, oriented along the $y$ axis with a period $\rho$. The total energy of this chain per one vortex, accounting for all mutual interactions, can be written as
\begin{gather}\label{E_chain}
    E_{tot/1v}(\rho)=E_v+\Delta E + \sum_{k=1}^{N_v-1}\frac{N_v-k}{N_v} \Big[ E_{v-v}(\rho k) + \Delta E_{int}(\rho k) \Big].
\end{gather}
The energy minimum $\text{min}_\rho E_{tot/1v}$ determines the equilibrium period $\rho_0$ of the bound state, and is shown in Fig.\ \ref{fig:SI_5}c for different $N_v$. One can clearly observe that the vortex attraction becomes stronger for larger clusters. For a moderate regime ($\ell\lesssim\lambda$) the system prefers the formation of groups with small VP number $N_v$. Namely, at $\ell/\lambda= 0.25$, the most favourable are three-VP clusters with a small internal size $\rho_0\ll\lambda$. This generally means that a long vortex chain is unstable with respect to the decay into small clusters separated by a large ($\gtrsim\lambda$) distance from each other (see Fig.\ \ref{fig:SI_5}a). Since the real material contains impurities and other types of mesoscopic inhomogeneities, the pinning of the VP and, consequently, the coexistence of clusters of different sizes are expected. Noteworthy, the same kind of effect has been predicted for completely different system of tilted vortices in thin anisotropic superconductors \cite{samokhvalovVortexClusters2010,samokhvalovAttractionPancake2012,buzdinVortexMolecules2013}.

We believe this to be a reasonable interpretation of the results observed in the experiment (see Fig.\ 4). One can understand the appearance of the magnetic bubbles as a ``fine-tuning effect": there is a specific range of $\ell\lesssim\lambda$ (controlled by the temperature) at which the stable multi-VP structures exist. At $\ell\gg\lambda$ the polaron effect is almost absent, while at $\ell\ll\lambda$ it may lead to the vortex generation instability, which requires additional investigation both from theory and experiment.

\revision{
\section*{Section S6: Modelling Vortex Profiles}

The magnetic force microscopy (MFM) vortex profiles were modelled using the Clem variational model\cite{clemSimpleModelVortex1975} modified to account for surface screening effects following an approach due to Kirtley \emph{et al}.\ \cite{kirtleyUpperLimitSpontaneous2007} assuming a variational coherence length $\xi_v (0) = 4.604$ nm. A Gorter-Casimir two fluid model temperature dependence has been assumed for the penetration depth, $\lambda(T) = \lambda(0) / \sqrt{1-(T/T_c)^4}$, with $\lambda(0) = 350$ nm\cite{stolyarovDomainMeissner2018}. The model provides the following description of the $z$-component of the $B$ field due to a vortex:
\begin{equation}
    B_z(x,y,z) = \frac{\phi_0}{2\pi\lambda}\int_0^\infty\frac{K_1(\sqrt{q^2+\lambda^{-2}\xi_v})}{(\sqrt{q^2+\lambda^{-2}}+q)K_1(\xi_v/\lambda)}J_0(q\sqrt{x^2+y^2})\exp(-qz)q\, dq \, ,
    \label{eq:clem_model}
\end{equation}
where $z$ is the scan height of the MFM sensor above the surface of the superconductor and has been treated here as a fit parameter. As discussed in the Methods section of the manuscript, the measured MFM frequency shifts ($\Delta_f = \Delta f_x + \Delta f_y$) are proportional to $\partial B_x / \partial x + \partial B_y / \partial y = - \partial B_z / \partial z$, and described in our model by
\begin{equation}
    -\dfrac{d B_z}{d z} (x,y,z) = \frac{\phi_0}{2\pi\lambda}\int_0^\infty\frac{K_1(\sqrt{q^2+\lambda^{-2}\xi_v})}{(\sqrt{q^2+\lambda^{-2}}+q)K_1(\xi_v/\lambda)}J_0(q\sqrt{x^2+y^2})\exp(-qz)q^2\, dq \, .
    \label{eq:dBzdz}
\end{equation}
In practice, due to the way the MFM data were recorded, the fits shown in the manuscript were made utilising three free parameters: the scan height $z$, a proportionality constant $\alpha$ and an arbitrary offset $\beta$, i.e.\ :
\begin{equation}
    \Delta f_x + \Delta f_y = \alpha\dfrac{d B_z}{d z}(x,y,z) + \beta \, .
    \label{eq:vortex_fit}
\end{equation}

\begin{figure}
    \centering
    \includegraphics[width=0.5\linewidth]{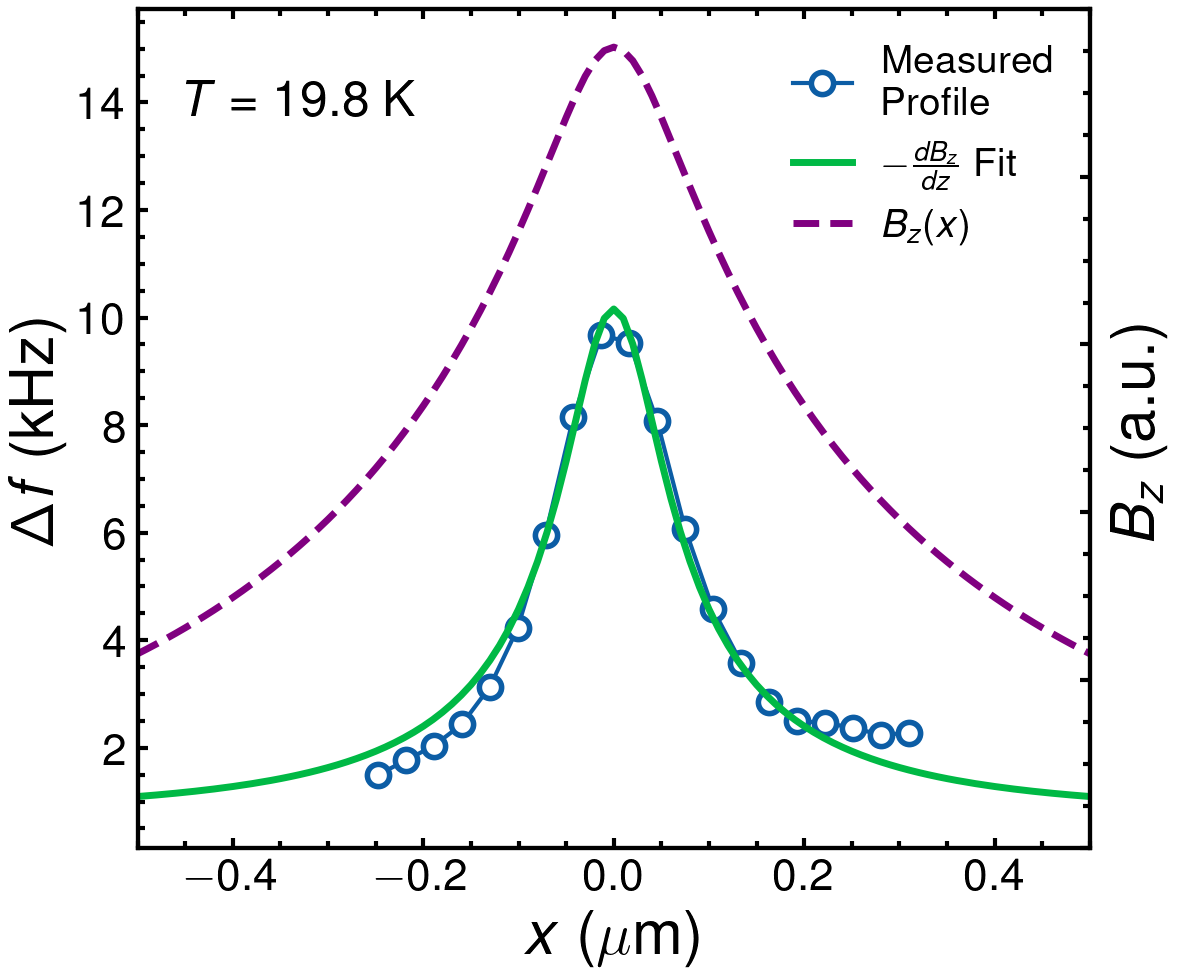}
    \caption{\revision{\textbf{Example MFM vortex profile and fit result.} A profile of a single vortex taken from an MFM measurement at $T = 19.8$ K (blue circles). The data are fitted to equation \ref{eq:vortex_fit} (green line) yielding a scan height of $z=49$ nm. The corresponding $B_z$ of the vortex is plotted in comparison (purple dashed line).}}
    \label{fig:SI_6}
\end{figure}

An example of the result of fitting a single vortex to \ref{eq:vortex_fit} at 19.8 K is shown in Fig.\ \ref{fig:SI_6} with fit parameters $z=49$ nm, $\alpha=0.072$ kHz/G and $\beta = 0.862$ kHz (assuming values of $\lambda(19.8 \, \mathrm{K}) = 462$ nm and $\xi_v(19.8 \, \mathrm{K}) = 10.5$ nm). Also plotted is an arbitrarily scaled version of $B_z(x,0,z)$ for the same scan height. Note how the MFM technique renders a much narrower version of the vortex than if one measured the vortex magnetic fields directly, in this case by almost a factor of three.

The profiles of double vortices that are bound together by vortex polarons (i.e.\ those shown in Fig.\ 4),  have been fitted to
\begin{equation}
    \Delta f_x + \Delta f_y = \alpha\left(\dfrac{d B_z}{d z}(x,y+\frac{w}{2},z) + \frac{dB_z}{dz}(x,y-\frac{w}{2},z)\right) + \beta \, ,
    \label{eq:double_vortex}
\end{equation}
where the additional fit parameter $w$ is the lateral separation between the two vortex cores.}
\bibliographystyle{apsrev}
\bibliography{SI-bib}